\documentclass[a4paper,fleqn,usenatbib]{mnras}

\usepackage{natbib}

\usepackage[T1]{fontenc}
\usepackage{ae,aecompl}

\usepackage{graphicx}	
\usepackage{amsmath}	
\usepackage{amssymb}	
\usepackage{color}
\usepackage[colorinlistoftodos]{todonotes}

\newcommand{\fu}{\textit{u}-band\,}
\newcommand{\fuS}{\textit{u}$^\ast$-band\,}
\newcommand{\flycfuv}{$f_{\mathrm{LyC}}/f_{\mathrm{UV}}$}
\newcommand{\flycfuvout}{$(f_{\mathrm{LyC}}/f_{\mathrm{UV}})^{\mathrm{out}}$}
\newcommand{\fesc}{$f_{\mathrm{esc}}$}

\title[LyC from AGNs at $z>3.3$]{Ionizing radiation from
AGNs at $z>3.3$ with the Subaru Hyper Suprime-Cam Survey and the CFHT Large Area
U-band Deep Survey (CLAUDS)}

\author[I. Iwata et al.]{
Ikuru Iwata,$^{1,2,3,4}$\thanks{E-mail: ikuru.iwata@nao.ac.jp}
Marcin Sawicki,$^4$\thanks{Canada Research Chair}
Akio K. Inoue,$^{5,6}$
Masayuki Akiyama,$^{7}$
\newauthor
Genoveva Micheva,$^8$
Toshihiro Kawaguchi,$^9$
Nobunari Kashikawa,$^{10}$
\newauthor
Stephen Gwyn,$^{11}$
Stephane Arnouts,$^{12}$
Jean Coupon,$^{13}$
Guillaume Desprez$^{13}$
\\
$^{1}$Subaru Telescope, National Astronomical Observatory of Japan, 650 North A'ohoku Place, Hilo, Hawaii 96720, USA\\
$^{2}$TMT Project, National Astronomical Observatory of Japan, 2-21-1, Osawa, Mitaka, Tokyo 181-8588, Japan\\
$^{3}$Department of Astronomical Science, The Graduate University for
Advanced Studies (Sokendai), 2-21-1, Osawa, Mitaka, Tokyo \\
181-8588, Japan\\
$^{4}$Department of Astronomy and Physics and Institute for
Computational Astrophysics, Saint Mary's University, 923 Robie Street, \\
Halifax, Nova Scotia B3H 3C3, Canada\\
$^{5}$ Department of Physics, School of Advanced Science and Engineering, Faculty of Science and Engineering, Waseda University, \\
3-4-1, Okubo, Shinjuku, Tokyo 169-8555, Japan\\
$^{6}$ Waseda Research Institute for Science and Engineering, Faculty of Science and Engineering, Waseda University, 3-4-1, \\
Okubo, Shinjuku, Tokyo 169-8555, Japan\\
$^{7}$ Astronomical Institute, Tohoku University, Aramaki, Aoba-ku, Sendai, Miyagi 980-8578, Japan\\
$^8$ Leibniz-Institute for Astrophysics Potsdam, An der Sternwarte 16, D-14482 Potsdam, Germany\\
$^9$ Department of Economics, Management and Information Science, Onomichi City University, Hisayamada 1600-2, Onomichi, \\
 Hiroshima 722-8506, Japan\\
$^{10}$ Department of Astronomy, Graduate School of Science, The University of Tokyo, 7-3-1 Hongo, Bunkyo-ku, Tokyo 113-0033, Japan\\
$^{11}$ NRC-Herzberg, 5071 West Saanich Road, Victoria, British Columbia V9E 2E7, Canada\\
$^{12}$ Aix Marseille Universit\'e, CNRS, Laboratoire d'Astrophysique de Marseille, UMR 7326, F-13388 Marseille, France\\
$^{13}$ Astronomy Department, University of Geneva, Chemin d'Ecogia 16, CH-1290 Versoix, Switzerland
}

\date{Accepted 2021 September 20. Received 2021 September 14; in original form 2021 January 26}

\pubyear{2021}

\usepackage{lscape}	
\begin{document}
\label{firstpage}
\pagerange{\pageref{firstpage}--\pageref{lastpage}}
\maketitle

\begin{abstract}
We use deep and wide imaging data from the CFHT Large Area U-band Deep
 Survey (CLAUDS) and the Hyper Suprime-Cam Subaru Strategic Program
 (HSC-SSP) to constrain the ionizing radiation (Lyman Continuum; LyC)
 escape fraction from AGNs at $z \sim 3 - 4$. For 94 AGNs with
 spectroscopic redshifts at $3.3 < z < 4.0$, we use their $U$-band /
 $i$-band flux ratios to estimate LyC transmission of individual
 AGNs. The distribution of their LyC transmission shows values lower
 than the range of LyC transmission values for IGM of the same redshift
 range, which suggests that LyC escape fraction of AGNs at $z>3.3$ is
 considerably lower than unity in most cases. We do not find any trend
 in LyC transmission values depending on their UV luminosities. Based on
 the photometry of stacked images we find the average flux ratio of LyC
 and non-ionizing UV photons escaping from the objects
 $(f_{\mathrm{LyC}}/f_{\mathrm{UV}})^{\mathrm{out}} = 0.182 \pm 0.043$
 for AGNs at $3.3<z<3.6$, which corresponds to LyC escape fraction
 $f_{\mathrm{esc}} = 0.303 \pm 0.072$ if we assume a fiducial intrinsic
 SED of AGN. 
 Based on the estimated LyC escape fraction and the UV luminosity
 function of AGNs, we argue that UV-selected AGNs' contribution to the
 LyC emissivity at the epoch is minor, although the size of their
 contribution largely depends on the shape of the UV luminosity
 function.
\end{abstract}




\section{Introduction}

Active Galactic Nuclei (AGNs) and massive stars in star-forming galaxies
are the two primary sources of hydrogen-ionizing radiation (Lyman
Continuum; LyC hereafter) in the Universe. 
Understanding the relative contributions from these two
populations to the ionizing photon budget over cosmic time is deeply
connected to our understanding of AGN and star-formation activity in
galaxies at different epochs.

In light of measurements of the faint-end galaxy UV luminosity function
(UVLF) at high redshift, it has been suggested that faint galaxies are
the population primarily responsible for reionizing the Universe
\citep[e.g.,][]{Inoue2006, Robertson2013, Dressler2015, Finkelstein2015,
Ishigaki2015}. However, maintaining reionization is not easy even with
the large number of faint galaxies that are observed: even if the UVLF
is integrated beyond the current observing limits, relatively high
ionizing photon escape fractions, $f_{\mathrm{esc}} = 10-20$\%, are
needed to keep intergalactic space ionized.  In contrast,  direct
constraints on \fesc\ for star-forming galaxies at $z\sim 3$
infer relatively low average values of $\lesssim$10\% 
\citep[e.g.,][]{Steidel2018, Iwata2019}. 
This tension suggests that \fesc\ in galaxies may be
luminosity-dependent or may increase at higher redshift
\citep{Inoue2006}. Alternatively, an additional source of ionizing
photons may be required.

Several studies based on the observations of the quasar UVLF have
reported that the AGN contribution to the ionizing photon budget is
minor \citep{Willott2010, Onoue2017, Akiyama2018, Kulkarni2019}.  
However, 
\citet{Giallongo2015} argued that the steep faint-end slope of the AGN
UVLF they found for X-ray selected AGNs at $4<z<6.5$ 
would imply that AGN could provide
enough photons to keep the Universe ionized. This view is supported by
\citet{Boutsia2018} who argued that, based on spectroscopy of faint AGNs
in the COSMOS field, the number density of faint AGNs could be higher
than found by earlier studies, and that, consequently, AGNs could make a
substantial contribution to the ionizing photon budget 
(see also \citet{Giallongo2019}, \citet{Grazian2020} 
and \citet{Boutsia2021} for further
reports on high AGN space density at $z\gtrsim4$).
Clearly, an accurate determination of the AGN UVLF is critically
important to give us a definitive evaluation of the AGN contribution to
the ionizing photon budget. But another critical parameter to be
understood here is the ionizing radiation escape fraction for AGNs. 

Previous studies often assumed $f_{\mathrm{esc}}=1$ on the supposition
that ionizing photons emerging from the AGN can efficiently escape into
the intergalactic space. However, studies of \fesc\ for AGNs at $z=3-4$
based on direct measurement of their LyC have shown that this assumption
could be wrong \citep{Cristiani2016, Micheva2017a, Grazian2018, Romano2019, 
Smith2020}, although the number of
AGN used in these studies is still small, 
except \citet{Cristiani2016} and \citet{Romano2019} who
examined $f_{\mathrm{esc}}$ of large numbers of bright quasars from SDSS
at $z>3.6$.
Given the potential
importance of AGN in reionizing the Universe, it is therefore important
to better constrain the AGN \fesc\ and its dependence on AGN luminosity.

The goal of the present paper is to constrain $f_{\mathrm{esc}}$ 
based on a large sample of AGNs with a broad range of UV luminosity.
Here, we turn to the Deep layer of the HSC Subaru Strategic Program
\citep[HSC-SSP;][]{Aihara2018, Aihara2019} which we combine with very
deep $U$-band images from the Canada-France-Hawaii Telescope (CFHT)
Large Area U-band Deep Survey \citep[CLAUDS, ][]{Sawicki2019}. 
These two surveys cover $\sim$19~deg$^2$ with very deep $u/u^*+grizy$
imaging to an unprecedented combination of area and depth; they span
enough volume with sufficient sensitivity to contain a significant
population of high-$z$ AGNs with a wide range of luminosities (more than
2 orders of magnitude in flux at rest-frame 1400\AA).
Together with a large spectroscopic sample of 94 AGNs at $3.3<z<4.0$ 
that we assembled in these fields from the literature, the deep CLAUDS
$u/u^*$-band images allow us to measure the ionizing flux escaping from
each spectroscopically-confirmed AGN, while the HSC-SSP photometry at
longer wavelengths provides their non-ionizing UV luminosity.

Throughout this paper we use AB magnitudes and assume the ($\Omega_M$,
$\Omega_\Lambda$, $H_0$) = (0.3, 0.7, 70~km~s$^{-1}$~Mpc$^{-1}$)
cosmology.

\section{Data}
This study uses very deep observed-frame $U$-band\footnote{Depending on
the field, CLAUDS uses \fu and \fuS filters \citep{Sawicki2019}. We
refer these two bandpass filters collectively as $U$-band.} fluxes to
determine the amount of ionizing photons escaping from AGN located at
redshifts $z>3.3$.  Photometry at longer wavelengths is used to
determine the intrinsic AGN luminosity which is needed to convert the
escaping ionizing flux into the ionizing photon escape fraction.
Accurate redshifts are essential for the AGN to ensure that the $U$-band
contains only ionizing radiation.  For these reasons our data consist of
a sample of AGNs with spectroscopic redshifts (described in
Section~\ref{sec:sample}) and photometry from two very deep, overlapping
imaging surveys (Section~\ref{sec:photometric_data}).

\subsection{Sample AGNs}
\label{sec:sample}

We use the imaging data from CLAUDS and HSC-SSP deep survey which
consist of four independent fields (XMM-LSS, Extended-COSMOS, ELAIS-N1,
and DEEP2-3) with areas of 4--6 deg$^2$ each and 18.60 deg$^2$ in total
and with $Ugrizy$ imaging (Section~\ref{sec:photometric_data}). 
We compiled a list of AGNs in the fields from the literature in
the redshift range between $z$=3.3 and 4.0. Our list consists of 94 
AGNs and contains only AGNs with redshifts marked as highly reliable
based on detections of multiple emission lines.
We excluded AGNs which are only listed in the catalogue of PRIMUS
\citep{Coil2011, Cool2013} even if their redshifts in the catalogue are
between 3.3 and 4.0, as reliability of the redshift listed in the
catalogue is not high for high redshift objects.
In Tables~\ref{tab:sample_xmmlss}, \ref{tab:sample_cosmos}, and
\ref{tab:sample_deep23_eliasn1} we give the positions, spectroscopic
redshifts, absolute magnitudes at rest-frame 1450 \AA, 
type of AGN (broad-line AGN (`BLA') or narrow-line AGN (`NLA')) in
literature if available, references and their designations in the
references.
For the AGNs in the SXDS / XMM-LSS field, we only use those at redshift
larger than 3.4, as only \fuS images are available in this field and in
the \fuS image there would be significant ($>25$\%) contributions from
non-ionizing photons for objects at $z<3.4$.

\begin{landscape}
\begin{table}
 \centering
\caption{A list of sample AGNs in the SXDS / XMM-LSS field.}
\label{tab:sample_xmmlss}
\begin{tabular}{crrcccccccll}
\hline
\multicolumn{12}{c}{SXDS / XMM-LSS}\\
\hline
ID  &  R.A.(J2000)& Decl.(J2000) & Redshift & $M_\mathrm{1450}$ & $U-i$ & $t_\mathrm{Lyc}$ & Filter & Type & N/E$^a$ & Ref.$^b$ & Designation \\
\hline
S01 & $ 33.9333920$ & $ -4.9228250$ & 3.5120 & $-24.604$ & $2.62\pm0.02$ & $0.140\pm0.003$ & \textit{u}$^\ast$ & BLA & N & 01,01 & SXDS0016 \\
S03 & $ 34.0578500$ & $ -4.1408925$ & 3.5190 & $-24.082$ & $>5.01$       & $<-0.017$       & \textit{u}$^\ast$ & BLA & N & 02,02 & SDSS021613.88$-$040827.2 \\
S07 & $ 34.3439178$ & $ -5.3945621$ & 3.4220 & $-21.816$ & $2.38\pm0.06$ & $0.114\pm0.012$ & \textit{u}$^\ast$ & NLA & E & 01,01 & SXDS0422 \\
S10 & $ 34.3655919$ & $ -5.2889194$ & 3.9806 & $-24.092$ & $>6.46$       & $<-0.008$       & \textit{u}$^\ast$ & BLA & N & 03,01 & SXDS0459 \\
S11 & $ 34.3932860$ & $ -5.0873914$ & 3.9832 & $-22.367$ & $>4.70$       & $<0.017$       & \textit{u}$^\ast$ & BLA & N & 04,01 & VANDELS\_UDS\_199159 \\
S13 & $ 34.4466743$ & $ -5.3992259$ & 3.5110 & $-24.313$ & $1.99\pm0.00$ & $0.285\pm0.003$ & \textit{u}$^\ast$ &  -- & E & 05,-- & IRS\_AAOmega:AAO\_SMA\_047 \\
S14 & $ 34.6180760$ & $ -5.2645695$ & 3.8570 & $-21.409$ & $>3.85$       & $<0.049$       & \textit{u}$^\ast$ & BLA & E & 01,01 & SXDS0809 \\
S15 & $ 34.6307498$ & $ -4.7318160$ & 3.6990 & $-23.626$ & $>5.93$       & $<-0.007$       & \textit{u}$^\ast$ & BLA & N & 01,01 & SXDS0824 \\
S16 & $ 34.6410242$ & $ -5.2876816$ & 3.5530 & $-22.219$ & $3.20\pm0.06$ & $0.078\pm0.001$ & \textit{u}$^\ast$ & BLA & N & 01,01 & SXDS0835 \\
S19 & $ 34.8146314$ & $ -4.4679767$ & 3.7611 & $-24.070$ & $3.17\pm0.01$ & $0.102\pm0.001$ & \textit{u}$^\ast$ & BLA & N & 06,06 & VIPERS:123049002 \\
S20 & $ 34.9913131$ & $ -4.4809647$ & 3.6400 & $-25.455$ & $5.05\pm0.02$ & $0.003\pm0.001$ & \textit{u}$^\ast$ & BLA & N & 06,06 & VIPERS:124045929 \\
S22 & $ 35.2521967$ & $ -4.3907918$ & 3.7148 & $-26.084$ & $3.00\pm0.00$ & $0.119\pm0.001$ & \textit{u}$^\ast$ & BLA & N & 07,07 & N\_19\_46 \\
S24 & $ 35.4857567$ & $ -5.8635564$ & 3.8466 & $-24.812$ & $>5.75$       & $<-0.005$       & \textit{u}$^\ast$ & BLA & N & 07,07 & N\_66\_19 \\
S25 & $ 35.6517807$ & $ -3.6446562$ & 3.5333 & $-26.128$ & $6.87\pm0.06$ & $-0.027\pm0.004$ & \textit{u}$^\ast$ & BLA & N & 02,02 & SDSS022236.43$-$033840.7 \\
S30 & $ 35.7661834$ & $ -5.5874712$ & 3.6719 & $-23.317$ & $3.48\pm0.03$ & $0.069\pm0.001$ & \textit{u}$^\ast$ & BLA & N & 07,07 & N\_97\_11 \\
S31 & $ 35.8362607$ & $ -3.3067372$ & 3.8653 & $-26.510$ & $>7.90$       & $<-0.013$       & \textit{u}$^\ast$ & BLA & N & 07,07 & N\_11\_3 \\
S35 & $ 36.0558630$ & $ -5.4569240$ & 3.7790 & $-25.849$ & $>8.20$       & $<-0.014$       & \textit{u}$^\ast$ & BLA & N & 02,02 & SDSS022413.41$-$052724.8 \\
S37 & $ 36.2992517$ & $ -4.4614914$ & 3.4905 & $-23.798$ & $2.15\pm0.01$ & $0.227\pm0.004$ & \textit{u}$^\ast$ & BLA & N & 07,07 & N\_50\_45 \\
S38 & $ 36.3634943$ & $ -4.4420211$ & 3.8527 & $-24.330$ & $>6.49$       & $<-0.009$       & \textit{u}$^\ast$ & BLA & N & 08,08 & VVDSDEEP:020254576 \\
S40 & $ 36.4314219$ & $ -4.4762785$ & 3.4147 & $-25.151$ & $2.30\pm0.00$ & $0.125\pm0.012$ & \textit{u}$^\ast$ & BLA & N & 02,02 & SDSS022543.53$-$042834.5 \\
S41 & $ 36.4375630$ & $ -4.6266875$ & 3.5900 & $-23.752$ & $>6.07$       & $<-0.013$       & \textit{u}$^\ast$ &  -- & N & 05,-- & 2XLSSdJ022545.0$-$043736 \\
S42 & $ 36.4611674$ & $ -4.3617302$ & 3.8600 & $-23.675$ & $2.76\pm0.01$ & $0.158\pm0.001$ & \textit{u}$^\ast$ & BLA & N & 08,08 & VVDSDEEP:020290592 \\
S44 & $ 36.6285575$ & $ -4.2239797$ & 3.5700 & $-22.651$ & $>5.02$       & $<-0.004$       & \textit{u}$^\ast$ &  -- & N & 05,-- & 2XLSSdJ022630.8$-$041326 \\
S48 & $ 36.9244832$ & $ -4.6017493$ & 3.7400 & $-21.148$ & $>2.91$       & $<0.132$       & \textit{u}$^\ast$ &  -- & E & 05,-- & 2XLSSdJ022741.9$-$043605 \\
S49 & $ 36.9744979$ & $ -4.3889907$ & 3.6260 & $-22.356$ & $>4.16$       & $<0.027$       & \textit{u}$^\ast$ & BLA & N & 08,08 & VVDSDEEP:020277536 \\
S50 & $ 36.9775822$ & $ -4.7598451$ & 3.7407 & $-25.328$ & $>7.17$       & $<-0.013$       & \textit{u}$^\ast$ & BLA & N & 02,02 & SDSS022754.61$-$044535.5 \\
S52 & $ 37.5102874$ & $ -4.5221641$ & 3.6581 & $-24.732$ & $>5.64$       & $<-0.005$       & \textit{u}$^\ast$ & BLA & N & 07,07 & N\_105\_36 \\
\hline
\end{tabular}
\begin{flushleft}
\textit{Notes.}
$^a$: `N' stands for a nuclear-dominated AGN, while `E' stands for an extended AGN (see Section~\ref{subsec:extendedness}).\\
$^b$: References. The first reference is for redshift, and the second one is for AGN type.
01: \citet{Akiyama2015},
02: SDSS quasar catalog DR14 \citep{Paris2018},
03: He et al., in prep., 
04: VANDELS DR2 \citep{Pentericci2018},
05: M. Akiyama, private communication,
06: VIPERS PDR2 \citep{Cucciati2017},
07: \citet{Menzel2016},
08: VVDS \citep{LeFevre2013}.
\end{flushleft}
\end{table}
\end{landscape}

\begin{landscape}
\begin{table}
 \centering
\caption{A list of sample AGNs in the COSMOS field.}
\label{tab:sample_cosmos}
\begin{tabular}{crrcccccccll}
\hline
\multicolumn{12}{c}{COSMOS}\\
\hline
ID  &  R.A.(J2000)& Decl.(J2000) & Redshift & $M_\mathrm{1450}$ & $U-i$ & $t_\mathrm{Lyc}$ & Filter & Type & N/E$^a$ & Ref.$^b$ & Designation \\
\hline
C07 & $149.4729034$ & $  2.7933720$ & 3.6095 & $-24.127$ & $3.18\pm0.01$ & $0.116\pm0.000$ & \textit{u}        & BLA & N & 09,09 & COSMOS\_J095753.49+024736.1 \\
C08 & $149.5337247$ & $  1.8091995$ & 3.9860 & $-23.357$ & $>5.54$       & $<0.015$       & \textit{u}        &  -- & N & 10,-- & B18\_0664641 \\
C10 & $149.6258478$ & $  1.9812633$ & 3.9600 & $-20.957$ & $2.15\pm0.12$ & $0.339\pm0.013$ & \textit{u}        & BLA & E & 11,11 & L450105 \\
C11 & $149.6959527$ & $  2.6030866$ & 3.3030 & $-21.326$ & $>3.95$       & $<-0.007$       & \textit{u}        & NLA & N & 11,11 & lid\_1808 \\
C12 & $149.7554065$ & $  2.7385464$ & 3.5240 & $-22.169$ & $>4.55$       & $<0.032$       & \textit{u}        & BLA & E & 12,09 & COSMOS\_J095901.31+024418.8 \\
C13 & $149.7820781$ & $  2.4712961$ & 3.3260 & $-21.819$ & $>4.33$       & $<0.003$       & \textit{u}        & BLA & N & 11,11 & L772453 \\
C14 & $149.8432151$ & $  2.6590580$ & 3.7480 & $-22.137$ & $>4.64$       & $<0.032$       & \textit{u}        & BLA & N & 10,10 & B18\_1730531 \\
C15 & $149.8457518$ & $  2.4816180$ & 3.3600 & $-21.927$ & $>4.42$       & $<0.022$       & \textit{u}        & BLA & N & 13,13 & M12\_1511846 \\
C16 & $149.8458637$ & $  2.8604351$ & 3.6300 & $-22.878$ & $>5.13$       & $<0.019$       & \textit{u}        & BLA & N & 10,10 & B18\_1938843 \\
C17 & $149.8494778$ & $  2.8652090$ & 3.6860 & $-21.854$ & $>4.07$       & $<0.053$       & \textit{u}        & NLA & E & 11,11 & lid\_2189 \\
C18 & $149.8515634$ & $  2.2764026$ & 3.3710 & $-22.412$ & $2.45\pm0.03$ & $0.202\pm0.001$ & \textit{u}        & BLA & N & 12,12 & XMMC\_149.85099+2.27609 \\
C19 & $149.8696906$ & $  2.2940175$ & 3.3450 & $-23.675$ & $>6.24$       & $<-0.010$       & \textit{u}        & BLA & N & 12,12 & XMMC\_149.86981+2.29391 \\
C20 & $149.8792145$ & $  2.2258072$ & 3.6510 & $-23.017$ & $>5.42$       & $<0.015$       & \textit{u}        & BLA & N & 12,12 & XMMC\_149.87949+2.22567 \\
C21 & $149.8861225$ & $  2.2759378$ & 3.3350 & $-21.944$ & $>4.52$       & $<0.005$       & \textit{u}        & NLA & E & 11,11 & cid\_1134 \\
C22 & $149.8942870$ & $  2.4329414$ & 3.3600 & $-21.393$ & $>3.88$       & $<0.043$       & \textit{u}        & BLA & N & 14,13 & L747071 \\
C23 & $149.9173822$ & $  2.8820631$ & 3.3170 & $-21.920$ & $2.40\pm0.05$ & $0.178\pm0.006$ & \textit{u}        & BLA & N & 14,14 & L1040434 \\
C24 & $150.0043858$ & $  2.0388885$ & 3.4990 & $-24.064$ & $3.24\pm0.01$ & $0.104\pm0.000$ & \textit{u}        & BLA & N & 12,12 & XMMC\_150.00412+2.03904 \\
C25 & $150.0229091$ & $  1.5862527$ & 3.3430 & $-21.297$ & $2.49\pm0.08$ & $0.182\pm0.004$ & \textit{u}        & NLA & E & 11,11 & lid\_1244 \\
C26 & $150.0426727$ & $  1.8721154$ & 3.3600 & $-21.882$ & $>3.81$       & $<0.047$       & \textit{u}        & BLA & N & 13,13 & PRIMUS\_110535 \\
C27 & $150.0968272$ & $  2.0214452$ & 3.5460 & $-20.424$ & $>2.68$       & $<0.180$       & \textit{u}        & NLA & E & 15,15 & CCOS1505 \\
C28 & $150.1139556$ & $  2.6067261$ & 3.9490 & $-20.793$ & $>2.72$       & $<0.198$       & \textit{u}        & BLA & E & 14,14 & L862385 \\
C29 & $150.1164373$ & $  1.9638991$ & 3.4100 & $-21.686$ & $>4.01$       & $<0.050$       & \textit{u}        & BLA & N & 13,13 & M12\_0790476 \\
C30 & $150.1303590$ & $  2.4659748$ & 3.8650 & $-22.764$ & $>4.80$       & $<0.028$       & \textit{u}        & BLA & N & 14,13 & L768961 \\
C31 & $150.2088406$ & $  2.4819033$ & 3.3330 & $-25.774$ & $>8.15$       & $<-0.027$       & \textit{u}        & BLA & N & 12,12 & XMMC\_150.20888+2.48202 \\
C32 & $150.2089844$ & $  2.4384687$ & 3.7150 & $-24.470$ & $>6.64$       & $<0.005$       & \textit{u}        & BLA & N & 12,12 & XMMC\_150.20929+2.43844 \\
C33 & $150.2407920$ & $  2.6590184$ & 3.3560 & $-22.578$ & $2.56\pm0.02$ & $0.172\pm0.001$ & \textit{u}        & BLA & N & 12,09 & XMMC\_150.24087+2.65873 \\
C34 & $150.2518296$ & $  1.5535407$ & 3.7470 & $-21.494$ & $>3.31$       & $<0.108$       & \textit{u}        & BLA & N & 14,14 & L179154 \\
C35 & $150.2595269$ & $  2.3761549$ & 3.7170 & $-23.402$ & $4.66\pm0.10$ & $0.031\pm0.000$ & \textit{u}        & BLA & N & 11,13 & M12\_1208399 \\
C36 & $150.2630206$ & $  2.5208549$ & 3.7580 & $-21.376$ & $>2.91$       & $<0.158$       & \textit{u}        & BLA & N & 14,14 & C1432719 \\
C37 & $150.2671997$ & $  1.9096446$ & 3.8460 & $-20.376$ & $>2.74$       & $<0.189$       & \textit{u}        & BLA & E & 14,14 & L405213 \\
C38 & $150.2715836$ & $  1.6138443$ & 3.5120 & $-20.122$ & $>2.67$       & $<0.177$       & \textit{u}        & NLA & N & 11,11 & cid\_1656 \\
C39 & $150.2972445$ & $  2.1487816$ & 3.3280 & $-25.123$ & $4.48\pm0.01$ & $-0.001\pm0.007$ & \textit{u}        & BLA & N & 12,12 & XMMC\_150.29764+2.14830 \\
C40 & $150.3007581$ & $  2.3005428$ & 3.4980 & $-21.426$ & $>3.98$       & $<0.053$       & \textit{u}        & NLA & E & 15,15 & CCOS784 \\
C41 & $150.3025734$ & $  1.8520637$ & 3.8400 & $-21.044$ & $>3.42$       & $<0.101$       & \textit{u}        & BLA & N & 13,13 & L368476 \\
C42 & $150.3060182$ & $  1.7616025$ & 3.3100 & $-20.338$ & $>3.01$       & $<0.075$       & \textit{u}        & NLA & E & 11,11 & cid\_3293 \\
C43 & $150.3445675$ & $  1.6359393$ & 3.4820 & $-19.177$ & $>1.69$       & $<0.431$       & \textit{u}        & NLA & E & 11,11 & cid\_1672 \\
C44 & $150.3647089$ & $  2.1437853$ & 3.3280 & $-22.050$ & $>3.72$       & $<0.032$       & \textit{u}        & BLA & E & 14,12 & L554731 \\
C45 & $150.3835833$ & $  2.0747463$ & 3.8590 & $-20.999$ & $>3.19$       & $<0.125$       & \textit{u}        &  -- & E & 16,-- & CCOS879 \\
C46 & $150.3941052$ & $  2.7178277$ & 3.4870 & $-21.600$ & $1.55\pm0.03$ & $0.492\pm0.006$ & \textit{u}        & NLA & E & 11,11 & lid\_4112 \\
C47 & $150.4029080$ & $  1.8788719$ & 3.5710 & $-20.394$ & $1.84\pm0.08$ & $0.393\pm0.012$ & \textit{u}        & NLA & E & 11,11 & cid\_2949 \\
C49 & $150.4399178$ & $  2.7034886$ & 3.4650 & $-22.510$ & $2.79\pm0.03$ & $0.155\pm0.001$ & \textit{u}        & BLA & N & 09,09 & COSMOS\_J100145.58+024212.6 \\
\hline
\end{tabular}
\end{table}
\end{landscape}

\begin{landscape}
\begin{table}
 \centering
\contcaption{A list of sample AGNs in the COSMOS field.}
\begin{tabular}{crrcccccccll}
\hline
\multicolumn{12}{c}{COSMOS}\\
\hline
ID  &  R.A.(J2000)& Decl.(J2000) & Redshift & $M_\mathrm{1450}$ & $U-i$ & $t_\mathrm{Lyc}$ & Filter & Type & N/E$^a$ & Ref.$^b$ & Designation \\
\hline
C50 & $150.4549406$ & $  1.9673839$ & 3.4710 & $-21.548$ & $>3.96$       & $<0.053$       & \textit{u}        & BLA & N & 11,11 & L441487 \\
C51 & $150.5508002$ & $  2.6828885$ & 3.5600 & $-21.818$ & $>4.11$       & $<0.048$       & \textit{u}        & BLA & N & 13,13 & M12\_1628943 \\
C52 & $150.6384404$ & $  2.3913200$ & 3.6500 & $-23.195$ & $>5.04$       & $<0.021$       & \textit{u}        & BLA & N & 13,13 & M12\_1159815 \\
C53 & $150.7037829$ & $  2.3699723$ & 3.7490 & $-23.966$ & $3.28\pm0.02$ & $0.112\pm0.000$ & \textit{u}        & BLA & N & 12,12 & XMMC\_150.70394+2.36961 \\
C54 & $150.7170733$ & $  1.9301231$ & 3.5675 & $-22.732$ & $1.23\pm0.01$ & $0.694\pm0.003$ & \textit{u}        & BLA & E & 11,11 & L419634 \\
C55 & $150.7355581$ & $  2.1995513$ & 3.4970 & $-25.257$ & $3.14\pm0.00$ & $0.063\pm0.006$ & \textit{u}$^\ast$ & BLA & N & 11,12 & lid\_1710 \\
C56 & $150.7371767$ & $  2.7225658$ & 3.3020 & $-23.399$ & $2.29\pm0.01$ & $0.186\pm0.009$ & \textit{u}        & BLA & N & 12,09 & COSMOS\_J100256.92+024321.2 \\
C57 & $150.7822171$ & $  2.2850682$ & 3.6260 & $-23.993$ & $3.79\pm0.03$ & $0.046\pm0.001$ & \textit{u}$^\ast$ & BLA & N & 10,10 & B18\_0899256 \\
C59 & $150.8013187$ & $  1.6574845$ & 3.7720 & $-23.212$ & $>5.63$       & $<0.013$       & \textit{u}        &  -- & N & 10,-- & B18\_0247934 \\
C60 & $150.9112639$ & $  1.9448424$ & 3.6800 & $-24.963$ & $4.43\pm0.02$ & $0.038\pm0.000$ & \textit{u}        & BLA & N & 03,03 & VVDSWIDE:100359356 \\
C61 & $150.9430180$ & $  1.3197523$ & 3.5570 & $-24.265$ & $>6.63$       & $<0.005$       & \textit{u}        & BLA & N & 03,03 & VVDSWIDE:100105943 \\
C62 & $151.2614791$ & $  1.4073308$ & 3.4035 & $-25.349$ & $3.07\pm0.00$ & $0.117\pm0.000$ & \textit{u}        & BLA & N & 02,02 & SDSS100502.75+012426.3 \\
\hline
\end{tabular}
\begin{flushleft}
\textit{Notes.}
$^a$: `N' stands for a nuclear-dominated AGN, while `E' stands for an extended AGN (see Section~\ref{subsec:extendedness}).\\
$^b$: References. The first reference is for redshift, and the second one is for AGN type.
02: SDSS quasar catalog DR14 \citep{Paris2018},
03: VVDS \citet{LeFevre2013},
09: \citet{Trump2009},
10: \citet{Boutsia2018},
11: \citet{Marchesi2016},
12: \citet{Brusa2010},
13: \citet{Masters2012},
14: \citet{Hasinger2018},
15: \citet{Civano2012},
16: \citet{Vito2014}.
\end{flushleft}
\end{table}
\end{landscape}

\begin{landscape}
\begin{table}
 \centering
\caption{A list of sample AGNs in the ELIAS-N1 and the DEEP2-3 fields.}
\label{tab:sample_deep23_eliasn1}
\begin{tabular}{crrcccccccll}
\hline
\multicolumn{12}{c}{ELIAS-N1}\\
\hline
ID  &  R.A.(J2000)& Decl.(J2000) & Redshift & $M_\mathrm{1450}$ & $U-i$ & $t_\mathrm{Lyc}$ & Filter & Type & N/E$^a$ & Ref.$^b$ & Designation \\
\hline
E01 & $241.6987255$ & $ 54.5804423$ & 3.3370 & $-26.156$ & $2.86\pm0.00$ & $0.119\pm0.003$ & \textit{u}        & BLA & N & 02 & SDSS160647.69+543449.6 \\
E03 & $242.3921983$ & $ 55.0041942$ & 3.5080 & $-25.186$ & $2.35\pm0.00$ & $0.239\pm0.000$ & \textit{u}        & BLA & N & 02 & SDSS160934.12+550015.1 \\
E04 & $242.4495213$ & $ 55.4285512$ & 3.4050 & $-25.993$ & $2.06\pm0.00$ & $0.299\pm0.000$ & \textit{u}        & BLA & N & 02 & SDSS160947.88+552542.7 \\
E05 & $242.7810873$ & $ 53.7022729$ & 3.9290 & $-25.054$ & $>6.89$       & $<0.004$       & \textit{u}        & BLA & N & 02 & SDSS161107.45+534208.1 \\
E07 & $242.9301107$ & $ 55.5325680$ & 3.5830 & $-25.534$ & $>7.48$       & $<0.002$       & \textit{u}        & BLA & N & 02 & SDSS161143.22+553157.3 \\
\hline
\multicolumn{12}{c}{DEEP2-3}\\
\hline
D03 & $350.9705935$ & $ -0.6618659$ & 3.5440 & $-24.981$ & $>7.20$       & $<0.003$       & \textit{u}        & BLA & N & 02 & SDSS232352.94$-$003942.7 \\
D04 & $351.1413339$ & $ -0.9803717$ & 3.5850 & $-24.985$ & $5.86\pm0.08$ & $0.010\pm0.000$ & \textit{u}        & BLA & N & 02 & SDSS232433.92$-$005849.3 \\
D07 & $351.3451664$ & $ -0.4108362$ & 3.6590 & $-25.192$ & $4.67\pm0.02$ & $0.030\pm0.000$ & \textit{u}        & BLA & N & 02 & SDSS232522.84$-$002439.1 \\
D08 & $351.4894363$ & $ -0.9161884$ & 3.7940 & $-25.405$ & $3.54\pm0.01$ & $0.089\pm0.000$ & \textit{u}        & BLA & N & 02 & SDSS232557.46$-$005458.3 \\
D09 & $351.8248088$ & $  0.0960467$ & 3.6760 & $-25.784$ & $>7.92$       & $<0.002$       & \textit{u}        & BLA & N & 02 & SDSS232717.95+000545.7 \\
D10 & $352.2085031$ & $  0.6830890$ & 3.6370 & $-24.252$ & $>6.68$       & $<0.005$       & \textit{u}        & BLA & N & 02 & SDSS232850.03+004059.0 \\
D11 & $352.2921745$ & $  0.2272098$ & 3.3980 & $-24.814$ & $2.17\pm0.00$ & $0.269\pm0.000$ & \textit{u}        & BLA & N & 02 & SDSS232910.12+001337.9 \\
D12 & $352.7057906$ & $ -0.0659099$ & 3.4490 & $-24.868$ & $2.54\pm0.00$ & $0.194\pm0.000$ & \textit{u}        & BLA & N & 02 & SDSS233049.38$-$000357.2 \\
D13 & $352.7191885$ & $ -0.5862646$ & 3.3360 & $-23.945$ & $2.62\pm0.01$ & $0.153\pm0.004$ & \textit{u}        & BLA & N & 02 & SDSS233052.60$-$003510.6 \\
\hline
\end{tabular}
\begin{flushleft}
\textit{Notes.}
$^a$: `N' stands for a nuclear-dominated AGN, while `E' stands for an extended AGN (see Section~\ref{subsec:extendedness}).\\
$^b$: 02: SDSS quasar catalog DR14 \citep{Paris2018}.
\end{flushleft}
\end{table}
\end{landscape}

\begin{figure}
 \includegraphics[width=\columnwidth]{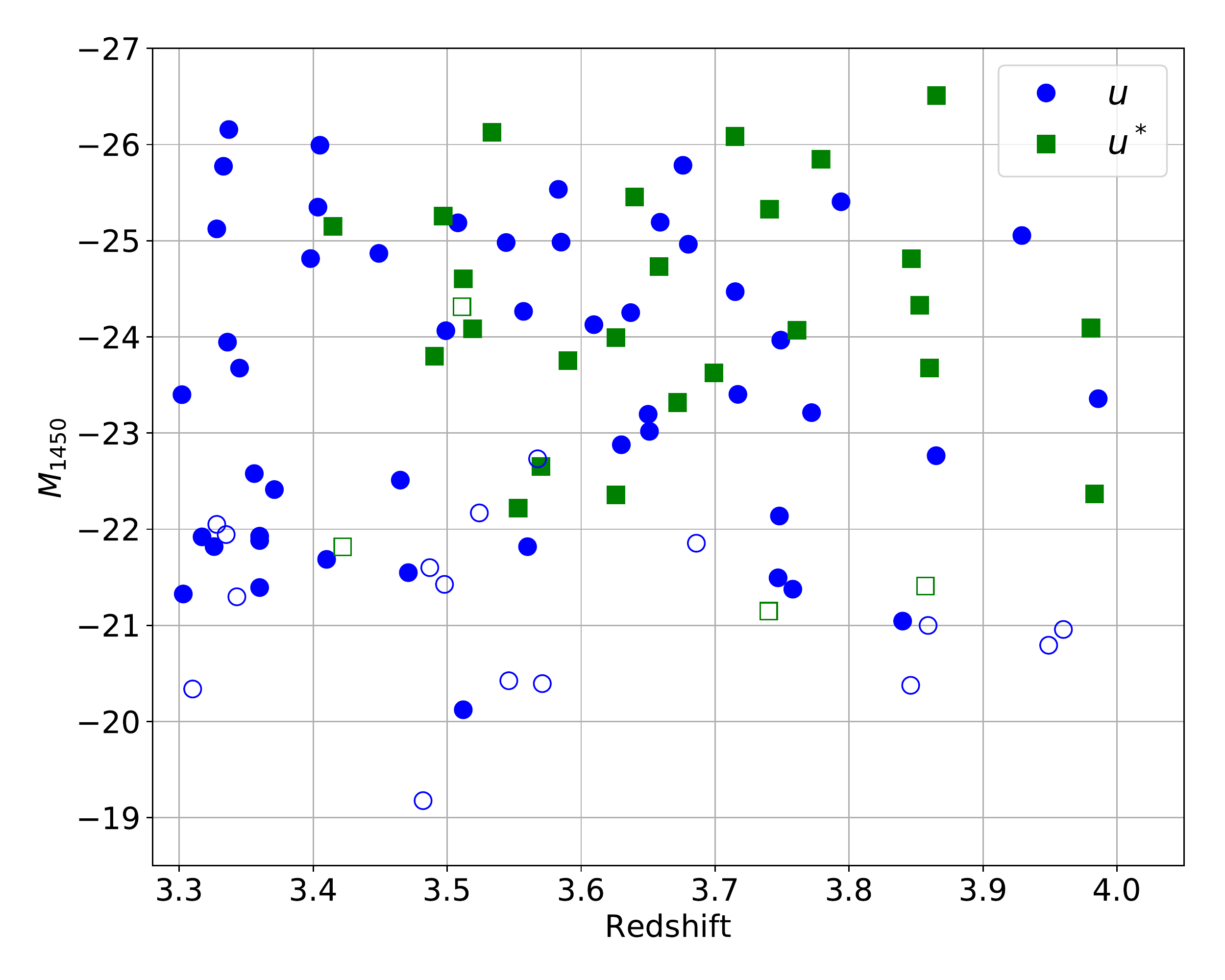}
 \caption{The redshift and absolute magnitude at rest-frame 1450 \AA\ 
 distribution of the sample AGNs. The blue points show the AGNs with \fu
 photometry, while the green squares show the AGNs with \fuS photometry.
 Filled symbols and open symbols represent nuclear-dominated and extended AGNs, respectively (see Section~\ref{subsec:extendedness}).
 }
 \label{fig:z_muv}
\end{figure}

Fig.~\ref{fig:z_muv} shows the distribution of redshifts and absolute 
magnitudes at rest-frame 1450\AA\ $M_\mathrm{1450}$ for our sample of
AGNs.
We calculate the $M_\mathrm{1450}$ magnitude of each AGN through a
\textit{K}-correction procedure that uses the synthetic colour
calculated from a fiducial quasar spectrum to infer the magnitude at
rest-frame 1450\AA\ from the observed \text{i}-band magnitude.
The fiducial spectrum comes from \citet{Lusso2015} and is based on the
\textit{HST} Wide Field Camera 3 spectra of 53 quasars at $z\simeq 2.4$
with correction of absorption by IGM. 
The \textit{K}-correction calculation is expressed as
\begin{equation}
M_\mathrm{1450} = i - DM + K,
\end{equation}
which we can expand as
\begin{equation}\label{eq:Kcorr}	
M_\mathrm{1450} = i + (m_\mathrm{1450} - i)_{\mathrm{synthetic}}  - 5 \mathrm{log}(d_L/10\mathrm{pc}) + 2.5 \mathrm{log} (1+z), 
\end{equation}
where the synthetic colour $(m_\mathrm{1450} - i)_{\mathrm{synthetic}}$
is the difference between magnitude at rest-frame 1450\AA\ and
\textit{i}-band magnitude for the fiducial AGN spectrum at the observed
redshift, and 
$d_L$ is a luminosity distance to the object. With cosmology fixed, the
only uncertainty in Equation~\ref{eq:Kcorr} is due to the synthetic
colour term, which could in principle be eliminated (as, for example, in
\citealt{Sawicki2006}) by using the observed $r$ magnitude (instead of
our $i$) as observed-frame $r$ corresponds closely to rest-frame
1450\AA\ at our redshifts. However, we use \textit{i}-band instead of
\textit{r}-band because at  $z>3.4$ the \textit{r}-band filter could
contain the Ly$\alpha$ emission line and, consequently, the flux
$r$-band flux density could strongly deviate from the continuum flux
density.  Nevertheless, while non-zero, $(m_{1450}-i)_{\rm synthetic}$
is still small and fairly constant, ranging over 0.15--0.22 mag
depending on the object's redshift.  The $M_{1450}$ values we calculated
are listed in Tables~\ref{tab:sample_xmmlss},
\ref{tab:sample_cosmos}, and
\ref{tab:sample_deep23_eliasn1}. 
The sample AGNs used in this study come from multiple catalogs in the
literature and contain both X-ray selected and optically selected
AGNs. We have compiled as many AGNs with reliable spectroscopic
redshifts as possible, and there is no uniform selection
criteria. Nevertheless, as Fig.~\ref{fig:z_muv} shows, 
there is no strong selection bias in terms of UV luminosity, and even in
the higher redshift range UV-faint AGNs are included in the sample.

\subsection{Photometric data}
\label{sec:photometric_data}

For photometry in $g$, $r$, $i$, $z$, and $y$-bands, we use the HSC SSP
S20A internal data release. We also tested photometry with the S18A data
release which was processed using an older version of the pipeline and
is identical to the second public data release
\citep[PDR2;][]{Aihara2019}. We confirmed that choice of the data
release does not alter our findings on the LyC transmission from the
sample AGNs significantly.
At shorter wavelengths, we use the \fu and \fuS observations provided by
CLAUDS \citep{Sawicki2019}.
The median depth of the data is $U$=27.1 AB (5$\sigma$ in 2\arcsec\
apertures), and there is a 1.36~deg$^2$ sub-area in the COSMOS field
with a median depth of $U$=27.7 AB  (5$\sigma$ in 2\arcsec\ apertures).
The median seeing size (FWHM) of the CLAUDS $U$-band data is 0\farcs92,
while for the HSC SSP Deep+UltraDeep data they are 0\farcs81, 0\farcs74,
0\farcs62, 0\farcs71 for $g$, $r$, $i$, $z$, $y$-bands, respectively
\citep{Aihara2019}.

\begin{figure}
 \includegraphics[width=\columnwidth]{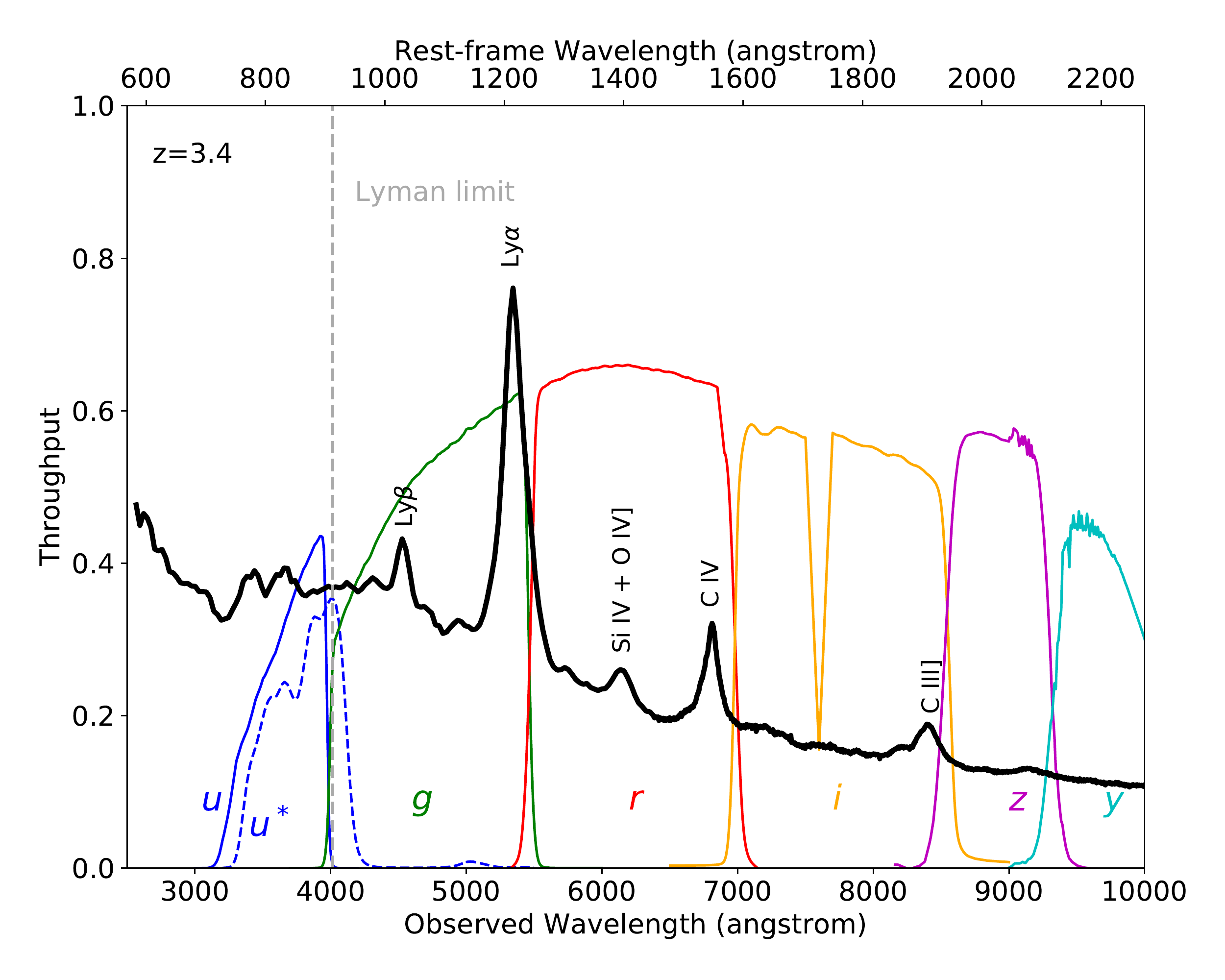}
 \caption{Transmission curves for filters for HSC and MegaCam 
 (solid line: \fu, dashed line: \fuS) used in this study. Filter
 transmissions include reflectance of the primary mirror, throughputs of
 camera optics, CCD quantum efficiency and 
 typical atmospheric transmission. 
 The \fuS transmission curve shows a small red leak at $\sim$5030\AA. 
 A stacked mean spectrum of quasar by \citet{Lusso2015} with redshift
 $z=3.4$ 
 (after correction of IGM attenuation)
 is shown with a thick black line. Lyman limit wavelength is
 indicated with a vertical dashed line.
 }
 \label{fig:filter_trans}
\end{figure}

In Fig.~\ref{fig:filter_trans}, throughput of the systems with these
bandpass filters are shown in the rest-frame wavelengths for sources at
$z=3.4$, along with the stacked quasar spectrum by
\citet{Lusso2015}. At $z>3.4$, \fu is free from non-ionizing photons,
while for \fuS there are some contributions by non-ionizing photons for
sources at $z>3.4$, because the filter transmission slope is shallower
than \fu and there is also a small leak around $\lambda = 5030$ \AA. In
Section~\ref{subsec:lyctrans} we will correct for such effects of
non-ionizing photons.
For the COSMOS field all the objects except two have both \fu and \fuS
images (the remaining two have only \fuS image). We use \fu for those
objects if both are available. For the objects in the ELIAS-N1 and
DEEP2-3 fields only \fu images are available, and in the SXDS/XMM-LSS
field only \fuS images are available.

\section{Analysis}
\label{sec:analysis}

\subsection{Photometry}

The coordinates of the sample AGNs are taken from their positions in the
HSC SSP database,
which is based on the $i$-band images.
For photometry in the HSC bands we use PSF model-based photometry 
\verb+[grizy]_psfflux_mag+ 
and its error \verb+[grizy]_psfflux_magerr+ in the forced catalog of the
database. For \textit{u} and \fuS images, we used {\sc photutils}
\citep{larry_bradley_2019_2533376} to obtain 1\farcs5-diameter aperture
photometry. 
Uncertainties were estimated with the help of artificial point sources
inserted into the images. The procedure is described in 
Appendix~\ref{appendix:mag_error}.
If an object's flux density is more than three times higher than the
1$\sigma$ error of the image, we regard the object is detected in
\textit{u} or \fuS, and use PSF model-based photometry in the HSC SSP
database, which employs the same algorithm used for photometry in HSC
bandpass filters, as its $U$-band magnitude. Otherwise the object is
regarded as undetected in $U$-band, and 1\farcs5-diameter aperture
photometry value is used to set the 3$\sigma$ upper-limit of its flux
density. 
We visually inspected postage stamp images of the all sample AGNs and
found that there is no object which has significant detection in
$U$-band with a spatial offset relative to \textit{i}-band. 

Finally, foreground Galactic dust extinction corrections were applied to
all the photometric data.  Here we used the dust reddening map by
\citet{Schlegel1998}, with the calibration by \citet{Schlafly2011},
obtained from the NASA/IPAC IRSA web
service\footnote{https://irsa.ipac.caltech.edu/applications/DUST/}. 

\begin{figure}
 \includegraphics[width=\columnwidth]{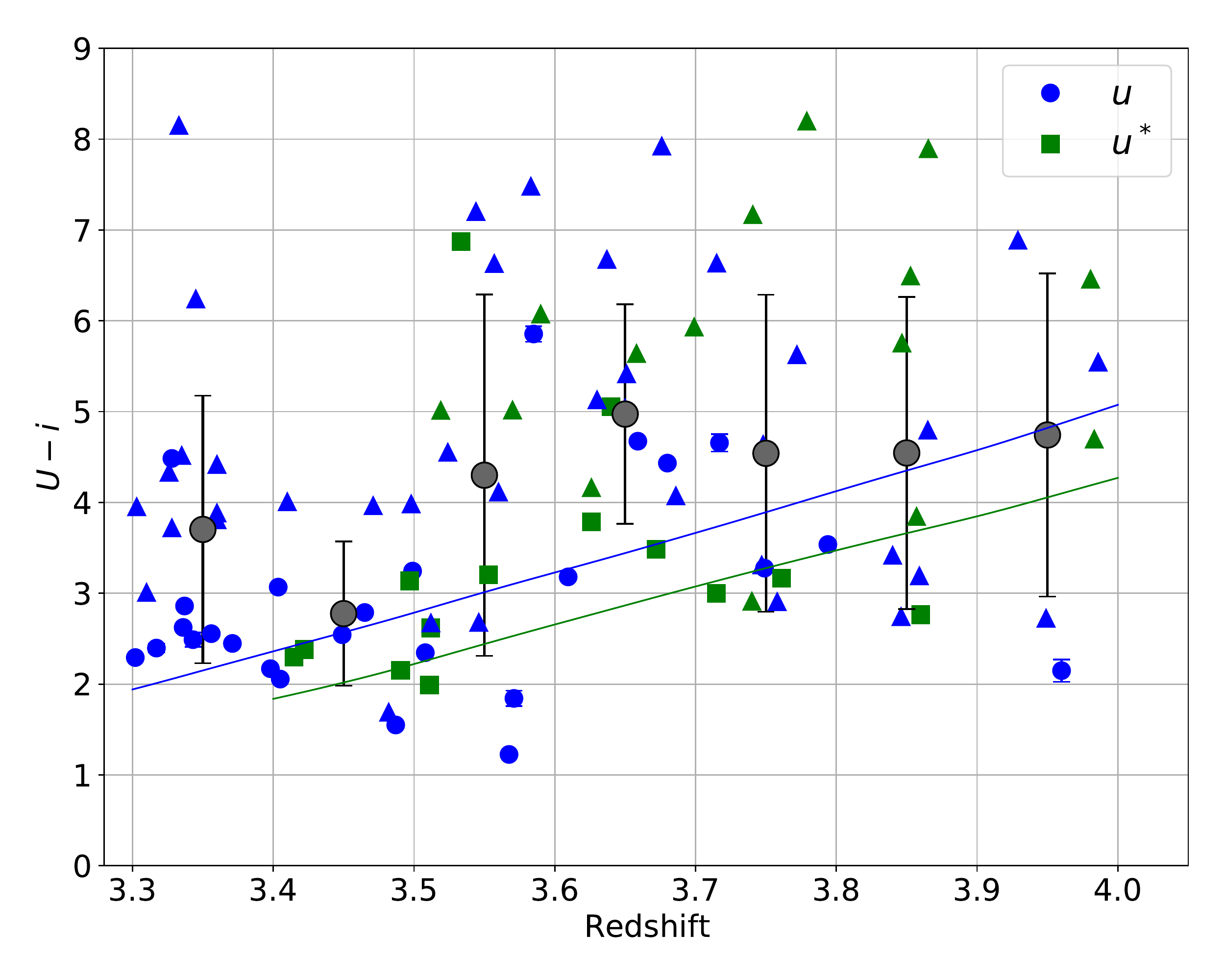}
 \caption{$U-i$ colours of the sample AGNs, plotted against their
 redshifts. Upper triangles represent lower limit $U-i$ colours for
 those without $U$-band detection. Larger black circles are average
 $U-i$ colours and their standard deviations for $\Delta z=0.1$
 bins. Blue and green solid lines show the expected colours using a
 fiducial intrinsic SED by \citet{Lusso2015}, average IGM attenuation
 \citep{Inoue2014}, and LyC escape fraction $f_{\mathrm{esc}} = 1.0$ for
 \fu and \fuS, respectively.
 }
 \label{fig:z_u-i}
\end{figure}

In Fig.~\ref{fig:z_u-i} $U-i$ colours of the sample AGNs are plotted
against their redshifts. 
The average $U-i$ colours of the sample AGNs are also plotted, in 0.1
redshift steps. When we calculate the average we simply use the lower
limit values for those not detected in $U$-band. Therefore these average
values should be taken as lower limits. 
In the figure we also show the expected $U-i$
colours if we use the stacked quasar spectrum by \citet{Lusso2015} as a
fiducial intrinsic SED, apply average IGM attenuation at the redshift
using a prescription given by \citet{Inoue2014}, and assume
$f_{\mathrm{esc}} = 1.0$. The expected colours increase along redshift,
due to increased average IGM attenuation. 
The observed $U-i$ colours of the sample AGNs roughly follow the trend of
the expected colours, although there are large dispersions in the
observed colours. As described in Section~\ref{sec:sample}, the sample
AGNs in this study come from multiple sources and no uniform selection
criterion was adopted.  
\citet{Prochaska2009} examined $u-g$ colours of SDSS quasars and found
that $u-g$ colours of $z\approx$3.5 quasars are systematically redder
than those at $z\approx$3.6. They argued that the SDSS colour-selection
criteria would introduce such selection bias. From Fig.~\ref{fig:z_u-i},
we do not see evidence for a bias of this type in terms of $U-i$ colours
for the sample AGNs in the present study.
The measured $U-i$ colours of
individual objects are given in Tables~\ref{tab:sample_xmmlss},
\ref{tab:sample_cosmos}, and \ref{tab:sample_deep23_eliasn1}.

\subsection{Selection of nuclear-dominated AGNs}
\label{subsec:extendedness}

\begin{figure}
 \includegraphics[width=\columnwidth]{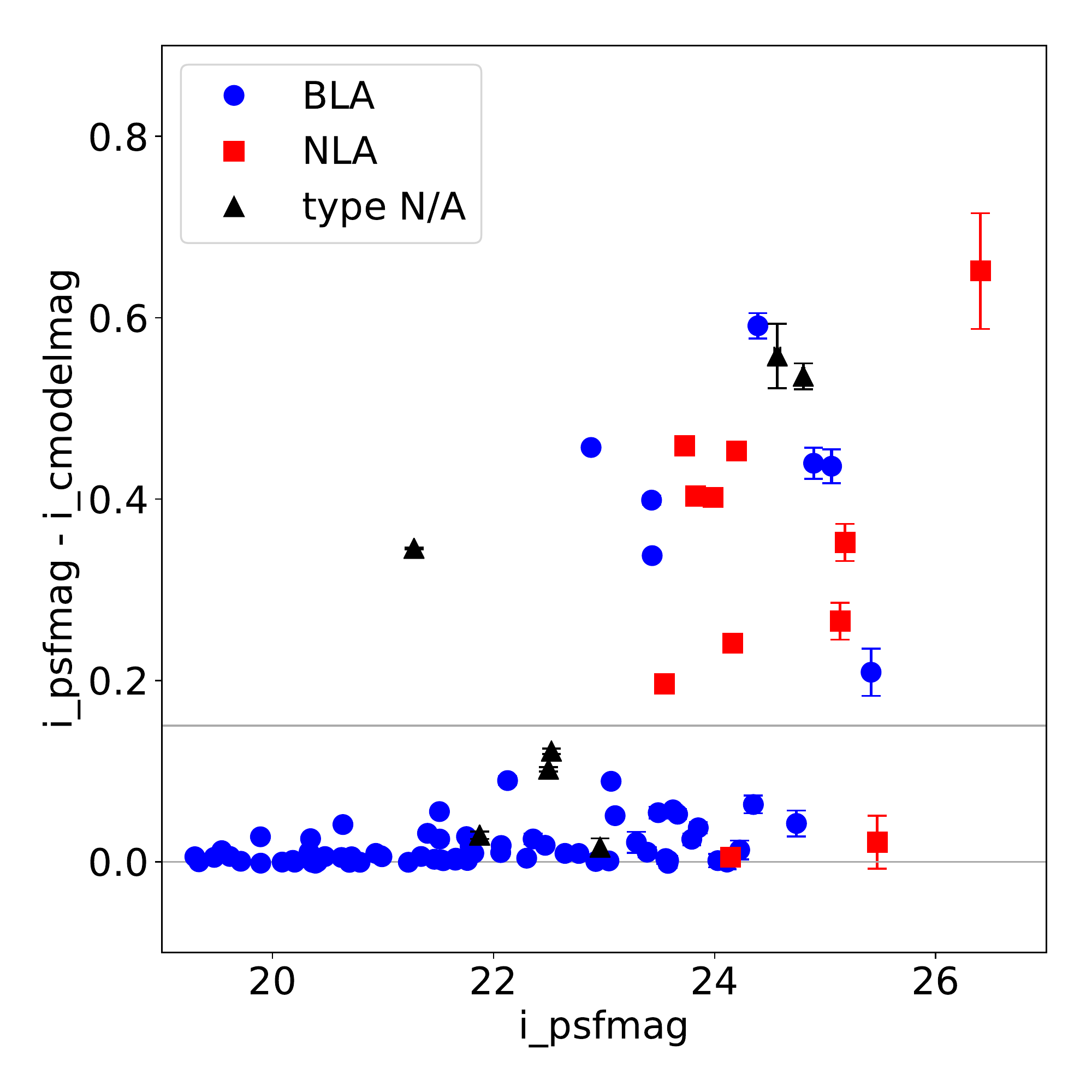}
 \caption{Magnitude difference between magnitude based on galaxy model
 fitting (CModel) and magnitude with PSF fitting against PSF fitting
 magnitude. 
 Blue circles and red squares represent AGNs classified in the reference
 literature as `Broad-line AGN' (BLA) and `Narrow-line AGN' (NLA),
 respectively, while black triangles are AGNs without BLA/NLA
 classification in the literature. A horizontal line at magnitude
 difference of 0.15 indicates the upper limit we adopted to select
 `nuclear-dominated' AGNs in this paper.
 }
 \label{fig:type}
\end{figure}

As described in Section \ref{subsec:lyctrans}, we need to assume an
intrinsic SED of AGN to estimate the LyC transmission and escape
fraction. For that purpose, we use the stacked quasar spectrum by
\citet{Lusso2015} as a template of intrinsic AGN SED. 
If the flux of an AGN is not dominated by its nucleus but its host
galaxy significantly contributes to its flux, we should use a different
SED to properly estimate the LyC transmission. However, it is difficult
to  estimate the intrinsic SED when the nuclear photons do not dominate
the SED. Thus we focus on sub-sample of AGNs dominated by nuclear
emission. 
In order to do so, we use the difference between magnitudes in $i$-band
measured by fitting PSF ($m_\mathrm{PSF}$) and those with photometry
using galaxy surface density profile model fitting
\citep[CModel;][]{Bosch2018}, $m_\mathrm{CModel}$. 
We use the following criterion to select nuclear-dominated AGNs:
\begin{equation}
    m_\mathrm{PSF} - m_\mathrm{CModel} < 0.15.
\end{equation}
This criterion means that flux density measured with galaxy model
fitting needs to be 
$\sim$15\% or less larger than the flux density
measured with PSF fitting.
Although this criterion is arbitrary, it
eliminates most of the AGNs which are classified as NLAs in literature
among the sample AGNs. As shown in Fig.~\ref{fig:type}, all but two
objects among 12 NLA AGNs have magnitude difference larger than 0.15,
and through visual inspection we find that the objects with magnitude
difference larger than 0.15 generally show extended morphologies in the
HSC images.

Among 94 sample AGNs, 74 objects satisfy this criterion. We refer this
subsample as `nuclear-dominated' AGNs and this subsample is used to
estimate LyC transmission and escape fraction.

\subsection{Estimation of LyC transmission}
\label{subsec:lyctrans}

We use \textit{u} and \fuS photometry to estimate LyC
emissivity and transmission of the target AGNs. 
Rest-frame wavelength ranges traced by these photometric data vary
depending on redshift of the source. For \fu the rest-frame wavelength
ranges traced by the filter with $>$50\% of the peak throughput are 800
-- 924 \AA\ and 688 -- 795\AA\ for an object at $z=3.3$ and $z=4.0$,
respectively, and for \fuS they are 789 -- 936 \AA\ and 694 -- 824 \AA\
for an object at $z=3.4$ and $z=4.0$, respectively.
For \fu at $z=3.3$ about 90\% of photons have $\lambda<$912\AA\, (for
both a template AGN SED and an SED that is flat in $f_\nu$), and at
$z\geq3.4$ all photons collected with this filter are LyC. On the other
hand, for \fuS, about 23\% of photons from a source at $z=3.4$
are non-ionizing UV photons, and  due to a small red leak in the
filter transmission around 5030 \AA (see Fig.~\ref{fig:filter_trans}),  
a few per cent of the photons collected with the filter is expected to
be non-ionizing photons even for a source at $z>3.6$. We need to correct
for contributions from such non-ionizing photons to calculate LyC
emissivity of the target AGNs.

LyC transmission for an object we want to know is defined as 
\begin{equation}
t_{\mathrm{LyC}} = f^\mathrm{obs}_\mathrm{LyC} / f^\mathrm{int}_\mathrm{LyC},
\label{eq:tLyC}
\end{equation}
where $f^\mathrm{obs}_\mathrm{LyC}$ and $f^\mathrm{int}_\mathrm{LyC}$
are the observed and intrinsic LyC flux densities, respectively. 
Note that $t_\mathrm{LyC}$ includes attenuation by both ISM and IGM. 
On the other hand, the escape fraction of LyC is the ratio of LyC
luminosity going out from an object to that generated within the object: 
\begin{equation}
 f_{\mathrm{esc}} = \frac{(L_{\mathrm{LyC}})^{\mathrm{out}}}{(L_{\mathrm{LyC}})^\mathrm{int}},
\end{equation} 
 and the relation between $t_{\mathrm{LyC}}$ and $f_{\mathrm{esc}}$ 
can be expressed as:
\begin{equation}
f_{\mathrm{esc}} =  t_{\mathrm{LyC}} \mathrm{exp}(\tau_{\mathrm{LyC}}^{\mathrm{IGM}}),
\end{equation}
where $\tau_{\mathrm{LyC}}^{\mathrm{IGM}}$ is IGM opacity for the
sightline of the object.
Because $\tau_{\mathrm{LyC}}^{\mathrm{IGM}}$ is unknown, we do not
derive $f_{\mathrm{esc}}$ for individual sources, but calculate 
only $t_{\mathrm{LyC}}$. We will estimate $f_{\mathrm{esc}}$ using
stacking analysis in Section~\ref{sec:stacking}.

Similar to $t_\mathrm{LyC}$,
transmission of non-ionizing UV photon is 
\begin{equation}
t_{\mathrm{UV}} = f^\mathrm{obs}_\mathrm{UV} / f^\mathrm{int}_\mathrm{UV}.
\end{equation}
If $U$-band contains non-ionizing photons,
the sum of $f^\mathrm{obs}_\mathrm{LyC}$ and
$f^\mathrm{obs}_\mathrm{UV}$ equals to the observed flux densities in
\textit{u} or \fuS, $f^\mathrm{obs}$. 
For $f^\mathrm{int}_\mathrm{LyC}$ and $f^\mathrm{int}_\mathrm{UV}$
we use  a mean $z\sim 2.4$ QSO spectrum by \citet{Lusso2015} as
an intrinsic SED. In Section~\ref{sec:sed_effect} we discuss how our
estimates of LyC transmission change if the assumed intrinsic SED is
changed.

For AGNs observed with \fu filter at $z\geq 3.4$, all photons collected 
by \fu filter are LyC photons, and we can simply use Eq. \ref{eq:tLyC} to 
calculate LyC transmission.
For AGNs observed with \fuS filter or those with \fu filter and at 
$z<3.4$, we need to estimate the amount of non-ionizing UV photons in the 
observation and subtract it to obtain $t_{\mathrm{LyC}}$:

\begin{equation}
    t_{\mathrm{LyC}} = \frac{f^{\mathrm{obs}}-f_{\mathrm{UV}}^{\mathrm{obs}}}{f_{\mathrm{LyC}}^{\mathrm{int}}}.
\label{eq:tLyC2}
\end{equation}

In order to estimate $f_{\mathrm{UV}}^{{\mathrm{obs}}}$, we need to take
the fluctuation of $t_{\mathrm{UV}}$ caused by the intervening IGM into
account. 
An absorber at a redshift close to the source produces both the LyC 
absorption and the Ly$\alpha$ absorption, and the degree of these 
absorption is determined by the opacity of the absorber. 
However, LyC photons travelling toward us can be also absorbed by an 
absorber at lower redshift due to Ly$\alpha$ absorption. This makes 
a variation in the values of $t_{\mathrm{LyC}}$ for sightlines toward 
sources at a redshift with a certain value of Ly$\alpha$ absorption, 
leading to an uncertainty in estimating $t_{\mathrm{UV}}$ and 
$t_{\mathrm{LyC}}$.
In order to consider the effect of such uncertainty, we use the results
of Monte Carlo simulation of IGM transmission \citep{Inoue2008} which
generate 10,000 sightlines for redshifts consistent with the H{\sc{i}}
cloud distribution defined analytically by \citet{Inoue2014}.  
First we use the mean UV transmission $t_{\mathrm{UV}}^0$ from 10,000
realizations of the sightlines with the redshift of a sample AGN to
make an initial estimate of LyC transmission, $t_{\mathrm{LyC}}^0$:

\begin{equation}
    t_{\mathrm{LyC}}^0 = \frac{f^{\mathrm{obs}}-f_{\mathrm{UV}}^{\mathrm{int}} \times t_{\mathrm{UV}}^0}{f_{\mathrm{LyC}}^{\mathrm{int}}}.
\end{equation}

Then we extract 1,000 instances from Monte Carlo realizations with 
$t_{\mathrm{LyC}}$ in the range of $t_{\mathrm{LyC}}^0 \pm
0.1$\footnote{Changing the range of $t_{\mathrm{LyC}}$ from Monte Carlo
realization does not affect the best estimate value of
$t_{\mathrm{LyC}}$ in most cases. By selecting sightlines with
$t_{\mathrm{LyC}}^0 \pm 0.15$ and $\pm0.20$ instead of
$t_{\mathrm{LyC}}^0 \pm 0.1$, the changes in resultant best estimate
values are less than 10\% for 69 and 60 objects out of the 74 sample
AGNs, respectively. The largest change in $t_{\mathrm{LyC}}$ is
$\sim30$\%.}. 
These simulated sightlines are used to get the distribution of 
$t_{\mathrm{UV}}$, and then to derive the mean value of 
$t_{\mathrm{LyC}}$, which is used as the best estimate of LyC
transmission of the object. The standard deviation of $t_{\mathrm{LyC}}$
values is added to the error estimate of $t_{\mathrm{LyC}}$ in addition
to the photometric errors.

In Tables~\ref{tab:sample_xmmlss}, \ref{tab:sample_cosmos}, and
\ref{tab:sample_deep23_eliasn1} $t_{\mathrm{LyC}}$ values calculated
with this procedure are provided.
Among 74 nuclear-dominated sample AGNs, 41 objects are undetected 
(i.e., have flux densities less than $3\sigma$) in \fu or \fuS images. 
For these objects, we estimate the upper limit of $t_{\mathrm{UV}}$ in
the same manner as detected sources but 
using $3\sigma$ upper limit flux density in \fu or \fuS, and if 
the estimated $f_{\mathrm{UV}}^{\mathrm{obs}}$, which equals to 
$f_{\mathrm{UV}}^{\mathrm{int}} \times t_{\mathrm{UV}}$, exceeds the
value of the $3\sigma$ upper limit flux density (i.e., upper limit of 
$t_{\mathrm{LyC}}$ is negative), we cannot put any constraint on 
$t_{\mathrm{LyC}}$ for the object. There are 14 such cases.

\subsection{Distribution of LyC transmission}

\begin{figure}
 \includegraphics[width=\columnwidth]{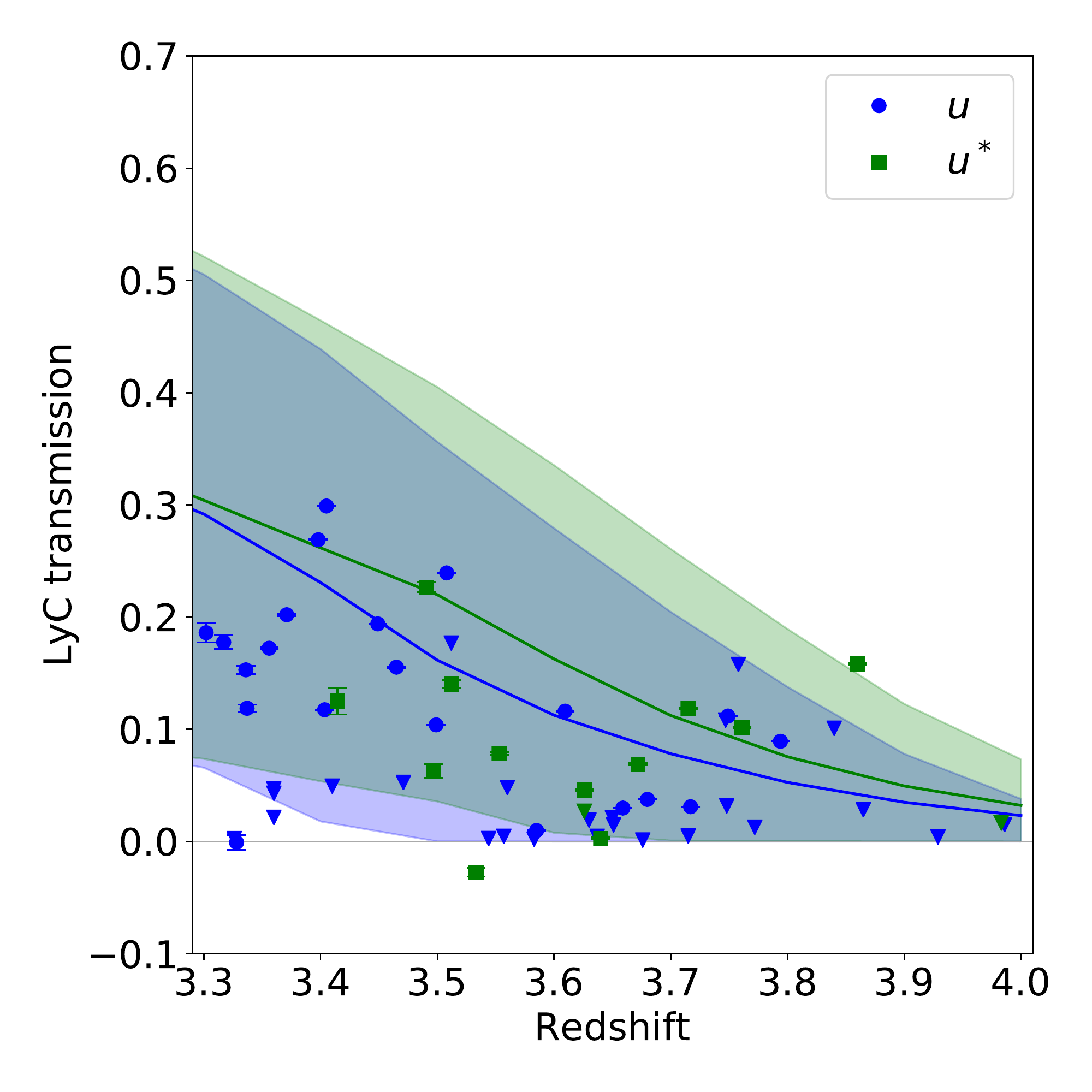}
 \caption{LyC transmission for the nuclear-dominated sample AGNs,
 plotted against their redshifts. 
 Blue circles are the values estimated using \fu photometry,
 and green squares are those using \fuS photometry. Downward triangles
 are 3$\sigma$ upper limits for the AGNs without detection in \fu or
 \fuS. The blue and green solid lines show the average IGM transmission
 for \fu and \fuS, respectively, from the Monte Carlo simulations, and
 shaded areas represent their 68\%-ile fluctuations.}
 \label{fig:tlyc_z}
\end{figure}

In Fig.~\ref{fig:tlyc_z} the distribution of LyC transmission
($t_\mathrm{LyC}$) of the 60 sample AGNs (14 objects with negative
$t_{\mathrm{LyC}}$ upper-limit among 74 objects are removed) 
is plotted against their redshifts.  
If \textit{u}/\fuS measured flux densities are
below the 3$\sigma$ detection limit then the $t_\mathrm{LyC}$ values
based on the 3$\sigma$ limits are presented as long as the upper limit
values are larger than the expected non-ionizing photon flux densities. 
Note that $t_\mathrm{LyC}$ includes attenuation by IGM. As the redshift
of a target AGN increases, the frequency of high IGM opacity along the
sightline becomes higher, as indicated by the solid lines and shaded 
areas in Fig.~\ref{fig:tlyc_z}. The distribution of $t_\mathrm{LyC}$
values for individual AGN also reflects this increase of IGM opacity
with redshift.  Note that the majority of the data points are under the
average IGM transmission (solid lines), which means that the average
$f_\mathrm{esc}$ of the sample AGNs is smaller than unity.

\begin{figure*}
 \includegraphics[width=\columnwidth]{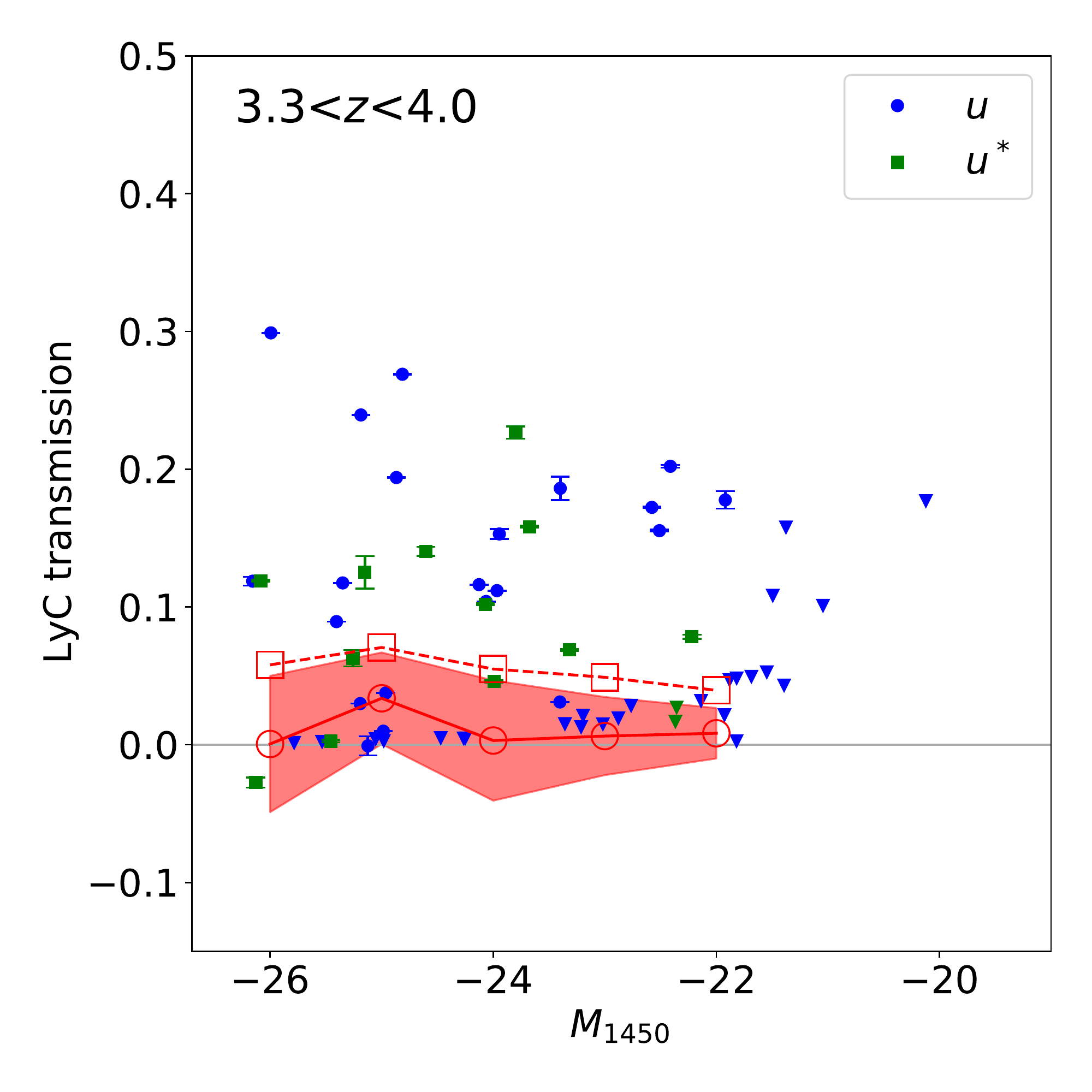}
 \includegraphics[width=\columnwidth]{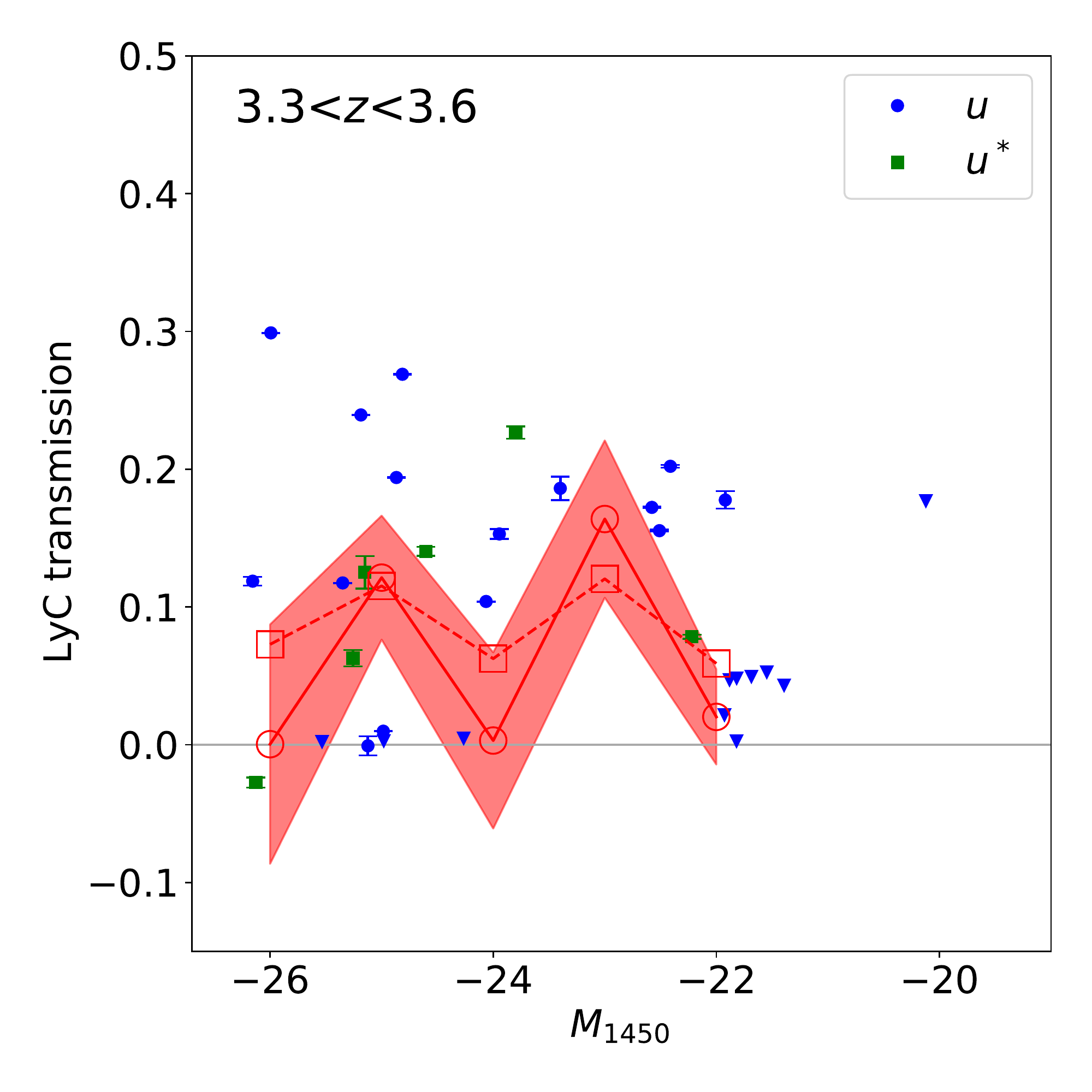}
 \caption{LyC transmission for the sample AGNs, plotted against their
 absolute magnitudes at rest-frame 1450\AA. 
 The calculated LyC transmission values ($t_\mathrm{LyC}$) based on
 $U$-band flux density measurements obtained through aperture photometry are
 plotted with filled circles (\fu) and filled squares (\fuS). 
 Downward triangles show 3$\sigma$ upper limits for the AGNs
 without detection in \fu or \fuS.
 (left) All the sample (nuclear-dominated) AGNs in the redshift
 range $3.3<z<4.0$ are plotted. The open circles and open squares
 indicate median and mean values in 0.5 magnitude bins, respectively,
 and the shaded area represents the variation of the  median values
 estimated by bootstrap resampling. 
 These values are calculated using LyC transmission 
 values based on aperture photometry, even for objects
 without $>$3$\sigma$ detection in $U$-band. 
 (right) Same as the left panel, but only for sample AGNs in the
 lower redshift range $3.3<z<3.6$.
}
 \label{fig:tlyc_muv}
\end{figure*}

To examine if there is a UV luminosity dependence on LyC transmission,
in Fig.~\ref{fig:tlyc_muv} $t_\mathrm{LyC}$ values are plotted against 
AGN's absolute magnitudes at rest-frame 1450\AA.
In the left panel of the figure we show the distribution of
$t_\mathrm{LyC}$ values of the nuclear-dominated AGNs at $3.3<z<4.0$.
In the right panel of the figure we also show the distribution of AGNs 
at $3.3 < z < 3.6$, a narrower redshift range where average IGM 
attenuation is relatively smaller than that for higher redshift. 
There are 38 AGNs in the redshift range.
In both panels we also plot the median and
mean values in 0.5 magnitude bins, as well as the variation of median
values estimated by bootstrap resampling. 
In calculation of median and mean values and variations, we use
$t_\mathrm{LyC}$ values based on 1\farcs5-diameter aperture photometry
even if the values are negative, while we use 3$\sigma$ upper-limits if
objects are not detected in $U$-band when plotting $t_\mathrm{LyC}$ of
individual AGNs.
No clear dependence on UV
luminosity is seen in neither panel. In the next subsection we further
examine if there is any UV luminosity dependence on LyC radiation from
AGNs, through stacking analysis.

\begin{figure*}
 \includegraphics[width=\columnwidth]{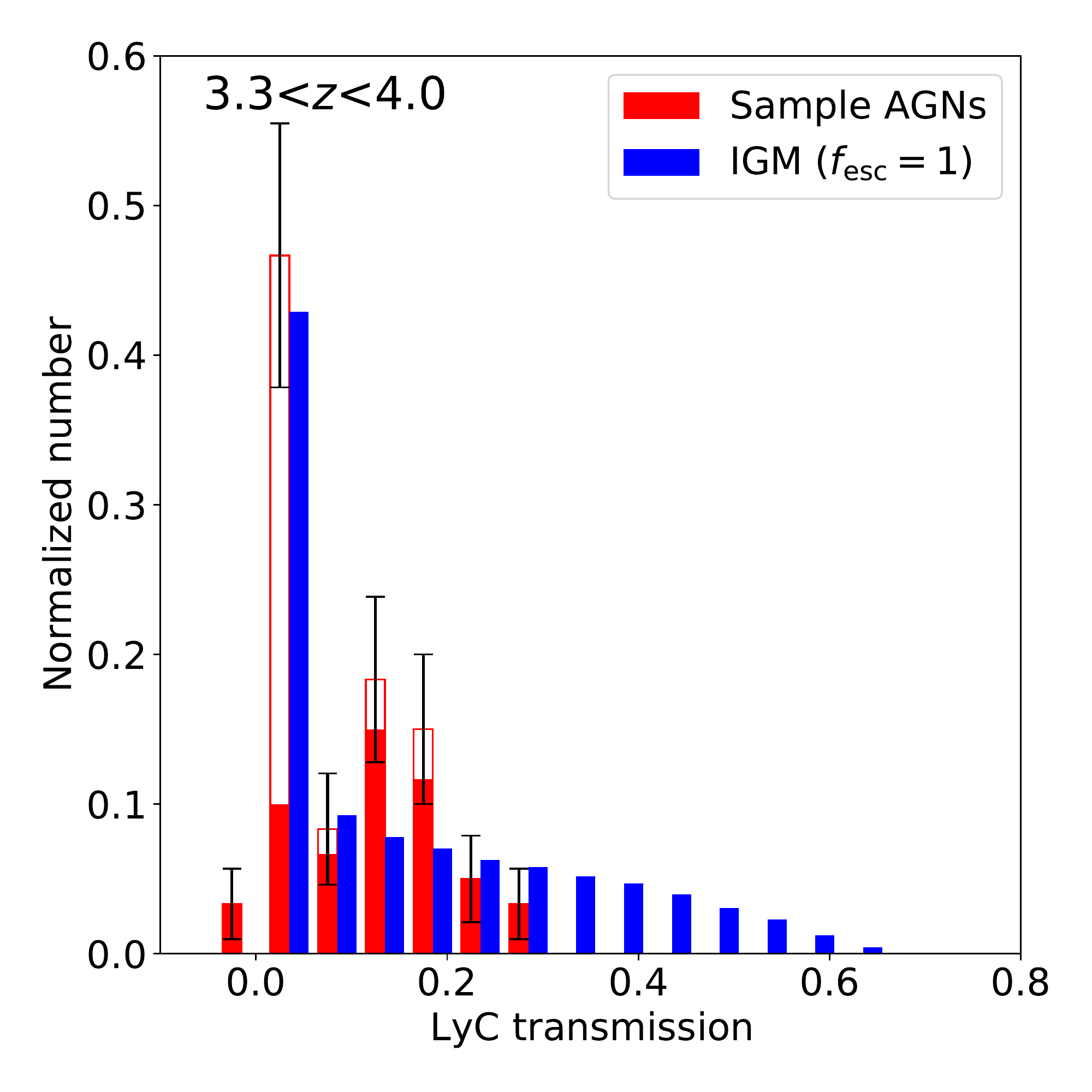}
 \includegraphics[width=\columnwidth]{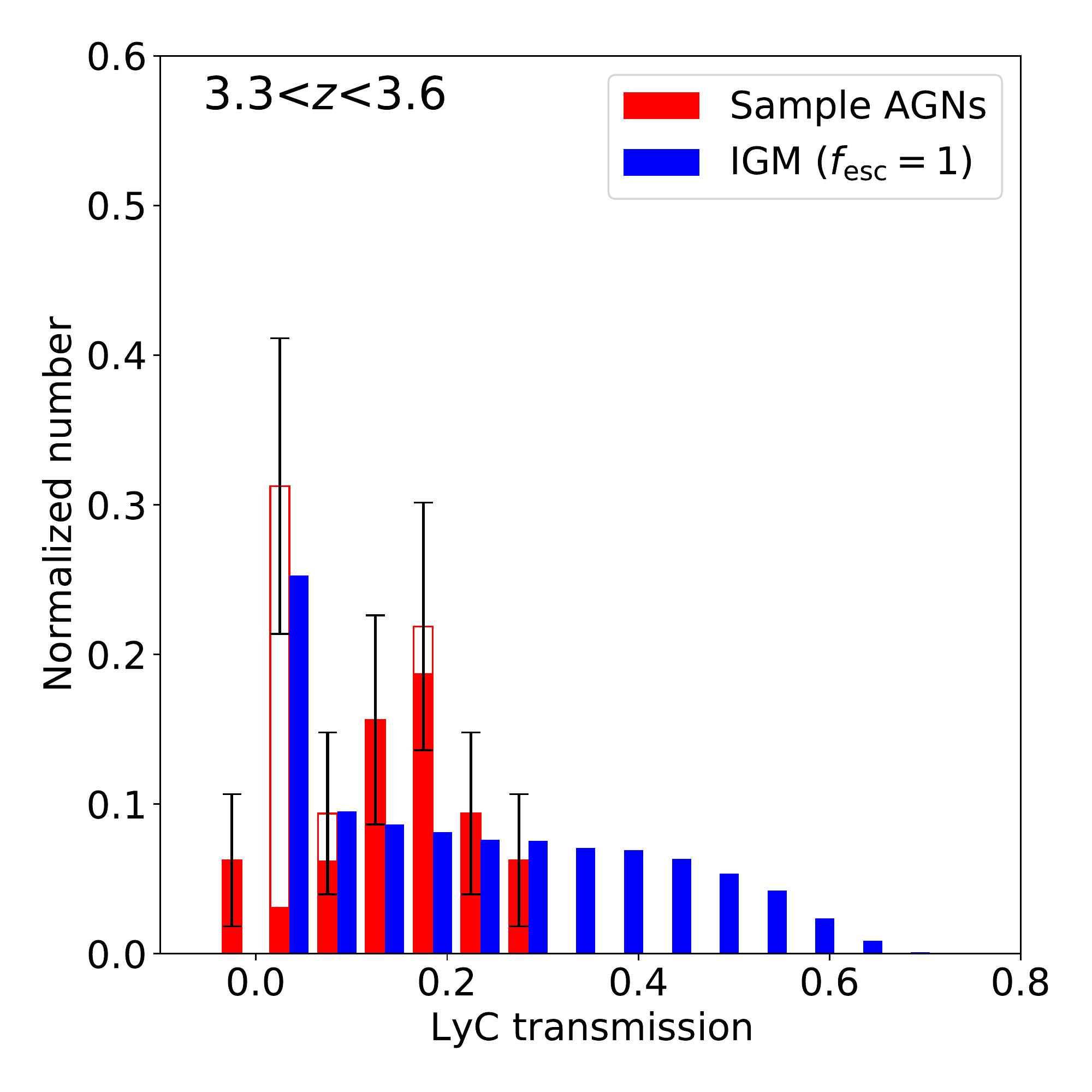}
 \caption{
 Histograms of LyC transmission for the sample AGNs.
 For both panels, normalized histogram of $t_{\mathrm{LyC}}$ for the
 sample (nuclear-dominated) AGNs and that of IGM LyC transmission at the
 same redshift range, which is derived from Monte Carlo realizations,
 are plotted. 
 Filled bars in $t_{\mathrm{LyC}}$ histograms represent those with
 $U$-band detections, while open bars are for those without $U$-band
 detections where we adopt 3$\sigma$ upper limits.
 Error bars for $t_{\mathrm{LyC}}$ of the sample AGNs show
 the sizes of Poisson error. 
 (left) All the sample AGNs in the redshift
 range $3.3<z<4.0$ are plotted.
 (right) Same as the left panel, but only for sample AGNs in the
 lower redshift range $3.3<z<3.6$.
}
 \label{fig:hist_tlyc}
\end{figure*}

In Fig.~\ref{fig:tlyc_muv} there appears to be a gap in the distribution
of $t_{\mathrm{LyC}}$ for both sample AGNs at $3.3<z<4.0$ and those at
$3.3<z<3.6$; a group of AGNs has $t_{\mathrm{LyC}} \simeq 0.0$ while
others have $t_{\mathrm{LyC}}$ around 0.1--0.3. 
In Fig.~\ref{fig:hist_tlyc}, we show the histograms of $t_{\mathrm{LyC}}$
of the sample AGNs. 
For those without $>3\sigma$ detection in $U$-band, we adopt 3$\sigma$
upper-limits of $t_{\mathrm{LyC}}$.
We also show the histograms of LyC transmission of IGM, which is
equivalent to the case when $f_\mathrm{esc}=1$. Those are generated
using the results of Monte Carlo realization of IGM distribution
calculated for 10,000 sightlines per redshift in 0.1 step between
$z=3.3$ and 4.0, and are combined with weighted average based on the
redshift distribution of the sample AGNs. 
From Fig.~\ref{fig:hist_tlyc} we see that the $t_{\mathrm{LyC}}$
distribution of the sample AGNs is more clustered towards lower values
than that of IGM, which is also clearly seen in
Fig.~\ref{fig:tlyc_z}. Such distribution indicates that LyC escape
fraction of the sample AGNs is considerably less than unity in most
cases. We also see that there is a second peak of $t_{\mathrm{LyC}}$ at
$\sim0.2$ in the $t_{\mathrm{LyC}}$ distribution of the sample AGNs. 
It should be reminded that the bimodality of LyC escape fraction has
been suggested in past studies on LyC escape of $z\sim3$ star-forming
galaxies \citep[e.g.,][]{Micheva2017b, Nakajima2020} 
and luminous quasars at $3.6<z<4.0$ \citep{Cristiani2016}.
The $t_{\mathrm{LyC}}$ distribution found in Fig.~\ref{fig:tlyc_muv} and
\ref{fig:hist_tlyc} may imply that LyC escape fraction distribution of
$3.3 < z < 4.0$ AGNs studied here 
also has a bimodal distribution. Unfortunately,
statistical significance is marginal due to a small number of the sample
AGNs, and there is a non-negligible possibility that such bimodal
distribution can be observed even if AGNs have a single $f_\mathrm{esc}$
value or their intrinsic $f_\mathrm{esc}$ distributes uniformly among
certain values between 0 and 1, due to the variance of IGM
attenuation. To confirm the bimodality of $f_\mathrm{esc}$ we need
larger sample of AGNs.

\subsection{Average LyC escape fraction from stacking analysis}
\label{sec:stacking}

We use the 38 objects in the redshift range $3.3 < z < 3.6$ among the
nuclear-dominated AGNs, including both with and without $U$-band
detections, for stacking analysis. First, $10\arcsec \times 10\arcsec$
subsections of \fu or \fuS  images centred at their $i$-band centroid
positions are extracted. We mask objects (except the sample AGNs)
detected by running  \textsc{SExtractor} version 2.19.5
\citep{Bertin1996} with 5 connected pixels above 2$\sigma$ as a
threshold. The local background value is estimated by getting a median
value of the count distribution for a $30\arcsec \times 30\arcsec$ area
around the object after iterations of 2$\sigma$ clipping, and the value
is subtracted from the image so that the median background value becomes
zero.
We then normalise the images with their $1\farcs5$ diameter aperture
flux density at rest-frame 1450\AA. 
The flux density at rest-frame 1450\AA\ of an object is derived 
from $m_\mathrm{1450}$ which is estimated in the way described in 
Section~\ref{sec:sample}.
We also correct for IGM attenuation by dividing the images
with mean IGM transmission with \textit{u} / \fuS at the object's
redshift, using the transmission formulation by \citet{Inoue2014}. 
The mean image is then generated by taking a mean of the array of counts
for each pixel. Measurement of the integrated counts with $1\farcs5$
diameter aperture gives the average \flycfuvout. 

\begin{figure}
 \includegraphics[width=\columnwidth]{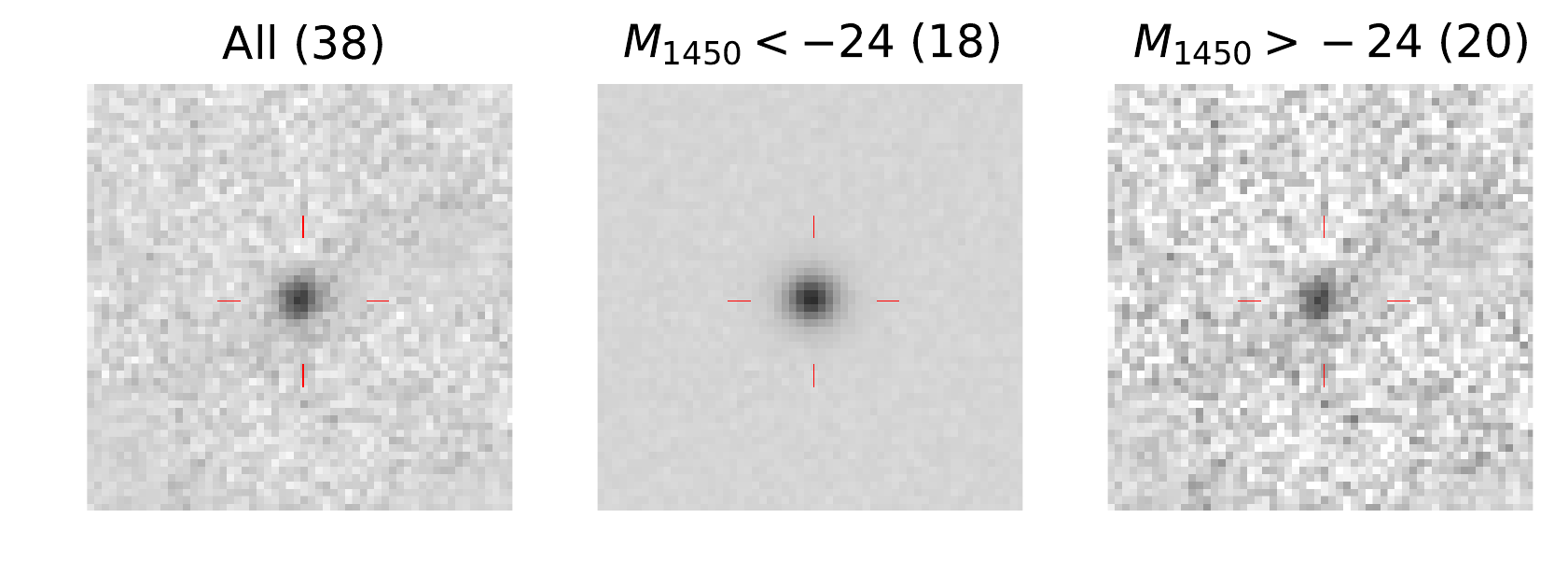}
 \caption{Stacked LyC images (observed-frame $U$-band). The images are
 normalized with UV flux density during stacking, and IGM attenuation
 has been corrected using the average value of IGM transmission at each
 object's redshift. 
 The image sizes are $10\arcsec \times 10\arcsec$.
~(Left) A stacked image using
 all of the 38 nuclear-dominated sample AGNs at $3.3<z<3.6$. 
~(Middle) A stacked image using 18 objects in the UV absolute
 magnitude $M_{\mathrm{1450}}<-24$.
~(Right) A stacked image using 20 objects in the UV absolute
 magnitude $M_{\mathrm{1450}}>-24$.
}
 \label{fig:stack}
\end{figure}

To search for indications of luminosity dependence on the LyC escape
fraction, we also generate stacked images of two subsamples based on
absolute UV magnitude.  
For this, we generate stacks with 18 objects and 20 objects with  
$M_{\mathrm{1450}} < -24$ and $M_{\mathrm{1450}} > -24$, respectively.
In Fig.~\ref{fig:stack} we show the stacked image with the full sample
of 38 objects, as well as images for two subgroups.

In the stacking procedure we assume the mean IGM attenuation for all
objects. This assumption would be valid to estimate the mean flux ratio
if the number of stacked objects is sufficiently large.  We estimate the
effect of fluctuation in IGM transmission by randomly selecting
sightlines from Monte Carlo IGM realisations and taking averages of
their IGM transmissions.  With 10,000 tests the standard deviation of
the average IGM transmission is $\sim$11\% for 38 sightlines and
15--17\% for 18 and 20 sightlines. These fluctuations are taken into
account when we estimate the error in stacking analysis to derive
average \flycfuvout. Additionally, we execute bootstrap resampling of
the images used for stacking and include the resulting scatter in the
error budget of our measurements. 
It should be also noted that we ignore any possible 
deviation of average IGM opacity around the 
sample AGNs from the average value determined from the observations 
of Lyman $\alpha$ absorbers \citep{Inoue2014}.
Strong ionizing radiation from AGNs may efficiently
ionize neutral hydrogen surrounding them and make their surrounding
environment more transparent to the ionizing radiation 
\citep[proximity effect; e.g.,][]{Scott2000}. 
On the other hand, AGNs may reside in the
peaks of matter distribution of the universe and H\textsc{i} column
density in the vicinity of the AGNs may be higher than the average. 
\cite{Prochaska2013} used close pairs of quasars at different redshifts
to find an excess of Ly$\alpha$ absorption in the transverse direction
of $z\sim 2$ quasars, with a correlation length of 12.5 Mpc for
optically thick ($N_{\mathrm{HI}}>10^{17.3}$ cm$^{-2}$)
absorbers.
Interestingly, \cite{Hennawi2007} argues that such excess of absorbers
is not observed in the the line of sight direction and claims
anisotropic ionizing radiation of quasars.
Although there are such possible complexities in the H{\sc i} gas
distribution in the vicinity of AGNs, we consider that it is fair to
adopt the mean IGM attenuation 
to estimate $f_{\mathrm{LyC}}$ in this study, because FWHMs of the
filter transmission are 530\AA\ and 647\AA\ for \fu and \fuS
respectively, which correspond to $\sim$550 Mpc in comoving scale for an
object at $z=3.4$. We measure LyC transmission along such a long line of
sight and therefore the effect of AGN environment in the H{\sc{i}}
column density would be insignificant.

Table~\ref{tab:stack} summarises the measured values of average
\flycfuvout\, for the 38 AGNs as well as bright and faint subgroups. 
Median stacked images are also generated, and their \flycfuvout\, 
values are also shown in the Table. 
The error sizes of bright and faint subsamples are
larger than those of the stacking of all objects, because the errors are
dominated by fluctuations in bootstrap tests and IGM transmission rather
than background noise, and a standard deviation with smaller number of
sample objects is larger.
We see no significant difference between \flycfuvout\, for the brighter
subgroup and that for the fainter one, which is also suggested by the
distribution of LyC transmission distribution of individual objects
shown in Fig.~\ref{fig:tlyc_muv}.
Here we do not take non-ionizing photons contained in \textit{u} / \fuS\
into account. Although the flux densities of non-ionizing photons from
individual objects are unknown, if we assume the intrinsic SED
\citep{Lusso2015} and use the mean IGM attenuation \citep{Inoue2014} at
the redshifts of the sample AGNs, the expected normalized counts of
non-ionizing photons in the stacked image is 0.013. The numbers of
$(f_{\mathrm{LyC}}/f_{\mathrm{UV}})^{\mathrm{out}}$ shown in
Table~\ref{tab:stack} would be smaller by $\approx$0.01 if we were able
to subtract non-ionizing photons from the \textit{u} / \fuS\  images.

The measured \flycfuvout\ value can be translated into LyC escape
fraction $f_{\mathrm{esc}}$ with an assumption of the intrinsic flux
ratio between LyC and UV (at rest-frame 1450\AA) by 
\begin{equation}
\label{eqn_fesc}
f_{\mathrm{esc}} = \frac{(f_{\mathrm{LyC}}/f_{\mathrm{UV}})^{\mathrm{out}}}{(L_{\mathrm{LyC}}/L_{\mathrm{UV}})^{\mathrm{int}}}.
\end{equation}

With the fiducial model SED, at the average redshift of the 38 sample
AGNs ($z=3.44$) the ratio between the flux traced with \fu and that at
rest-frame 1450 \AA\, is 0.601. The \fesc values using this
ratio as  $(L_{\mathrm{LyC}}/L_{\mathrm{UV}})^{\mathrm{int}}$ are also
listed in Table~\ref{tab:stack}.

We also examine how \flycfuvout\ varies if AGNs in different redshift
ranges are used for stacking. 
In Table~\ref{tab:stack} we show \flycfuvout\ and \fesc\ values for the
case with a narrower redshift range ($3.3<z<3.5$) and the case in which
the redshift range is shifted to higher redshift by $\Delta z=0.1$
($3.4<z<3.7$). The \flycfuvout\ value for AGNs at $3.3<z<3.5$ is higher
than the value of those at $3.3<z<3.6$, and 
\flycfuvout\ value for AGNs at $3.4<z<3.7$ is lower than the case with
$3.3<z<3.6$, although the differences are within standard deviations and
thus are statistically insignificant. Possible causes of these
differences would include the evolution of \fesc\ (smaller \fesc\ at
higher redshift) and additional IGM attenuation with respect to the IGM 
model by \citet{Inoue2014} at higher redshift or at shorter wavelength.
The size of the present AGN sample would not allow us to further
investigate the redshift evolution of \flycfuvout\ and \fesc, but the
readers should be aware of such uncertainties in \flycfuvout\ and \fesc\
values in this study. \flycfuvout\ does not depend on the shape of the
intrinsic SED, while the conversion from \flycfuv\ to \fesc\ uses
$(L_{\mathrm{LyC}}/L_{\mathrm{UV}})^{\mathrm{int}}$
(Equation~\ref{eqn_fesc}).
We will further discuss the effect of changes in
the assumed intrinsic spectrum in Section~\ref{sec:sed_effect}.

\begin{table*}
 \centering
\caption{Results of stacking analysis with AGNs.}
\label{tab:stack}
\begin{tabular}{ccccccc}
\hline
  &  Number of & \multicolumn{2}{c}{\flycfuvout$^a$} & &\multicolumn{2}{c}{$f_{\mathrm{esc}}^b$} \\
\cline{3-4} \cline{6-7}
  &  objects & Mean & Median & &Mean & Median \\
\hline
\multicolumn{7}{c}{$3.3<z<3.6$}\\
\hline
all                   & 38 & 0.182$\pm$0.043 & 0.163$\pm$0.057 & & 0.303$\pm$0.072 & 0.271$\pm$0.094 \\
$M_{\mathrm{1450}}<-24$ & 18 & 0.215$\pm$0.062 & 0.182$\pm$0.087 & & 0.358$\pm$0.104 & 0.303$\pm$0.145 \\
$M_{\mathrm{1450}}>-24$ & 20 & 0.152$\pm$0.061 & 0.144$\pm$0.068 & & 0.253$\pm$0.101 & 0.240$\pm$0.112 \\
\hline
\multicolumn{7}{c}{$3.3<z<3.5$}\\
\hline
all & 25 & 0.247$\pm$0.050 & 0.233$\pm$0.059 & & 0.410$\pm$0.084 & 0.388$\pm$0.099 \\
\hline
\multicolumn{7}{c}{$3.4<z<3.7$}\\
\hline
all & 37 & 0.142$\pm$0.041 & 0.095$\pm$0.046 & & 0.236$\pm$0.068 & 0.158$\pm$0.076 \\
\hline
\end{tabular}
\begin{flushleft}
\textit{Notes.}
$^a$: Correction to IGM attenuation is made with assumption of mean IGM
 attenuation by \citet{Inoue2014}.
$^b$: Assuming the intrinsic \flycfuv = 0.601 based on the AGN SED by
 \citet{Lusso2015}.
\end{flushleft}
\end{table*}

\section{Discussion}

\subsection{The effect of assumed intrinsic AGN SED}
\label{sec:sed_effect}

There is a large dispersion in the $\lambda \lesssim 1200$\AA\ continuum
slopes of AGNs  reported in the literature. One cause of such
differences is the difficulty in correcting for IGM absorption. However,
there may also be intrinsic differences in the continuum slopes
depending on the redshift and luminosity of the sample AGNs  
\citep[see detailed discussion in][]{Lusso2015}.
Here we examine how our estimates of LyC transmission,
$t_{\mathrm{LyC}}$, and LyC escape fraction, $f_{\mathrm{esc}}$, change
when we alter our assumptions about the intrinsic AGN SED. 

\begin{figure}
 \includegraphics[width=\columnwidth]{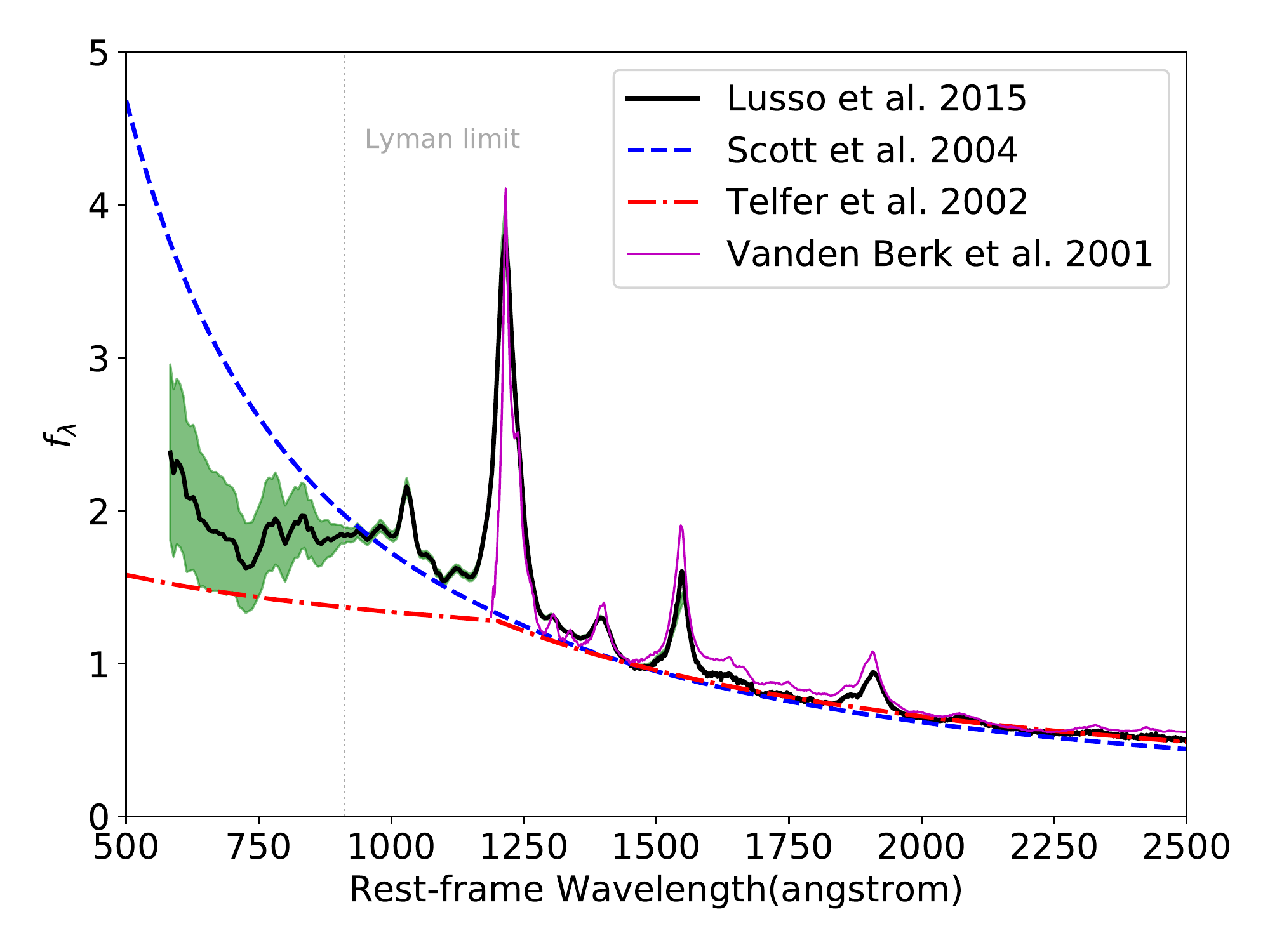}
 \caption{Model AGN SEDs normalised at rest-frame 1450\AA. 
 The thick black line and the green shaded area are the $z\sim 2.4$ quasar
 SED by \citet{Lusso2015} which is used as the fiducial SED in this
 study and its uncertainties. 
 The blue dashed line and red dot-dashed line are continuum 
 slopes fitted to the quasar composite spectra reported by
 \citet{Scott2004} and \citet{Telfer2002}, respectively. 
 The thin magenta line is a composite quasar spectrum based on SDSS 
 spectroscopy reported by \citet{VandenBerk2001} (only shown for
 wavelength $>$1200\AA).
}
 \label{fig:modelsed}
\end{figure}

In addition to the baseline AGN SED that we used in
Section~\ref{sec:analysis}, we now consider two continuum slopes from
the literature that differ significantly from our baseline SED model and
from each other at rest-frame wavelength $\lambda \lesssim 1200$\AA. 
One is the slope reported by \citet{Telfer2002}, which is based on
\textit{HST} UV spectra of 184 quasars at $z>0.33$ (most of their sample
AGNs are at $z<2.5$). The fitted continuum slope for their composite
spectrum is $\alpha = -1.76$ ($f_\nu \propto \nu^\alpha$) for 
500\AA$< \lambda \lesssim 1200$\AA\, and $\alpha = -0.69$ 
for $\lambda \gtrsim 1200$\AA. The other is the slope reported by
\citet{Scott2004}, which is based on \textit{FUSE} spectra of $\sim$100
AGNs at $z<1$. This sample contains AGNs at lower redshifts than those in the
samples used by \citet{Telfer2002} and \citet{Lusso2015}; moreover, less
luminous Seyfert 1 AGNs are included in this sample, while the samples of
\citet{Telfer2002} and \citet{Lusso2015} contain only quasars.
At $\lambda \lesssim 1200$\AA\ the  continuum slope for this sample is
$\alpha = -0.56$, which is much steeper than the slope reported by
\citet{Telfer2002}; at $\lambda \gtrsim 1200$\AA\ the slope is $-0.83$,
which is similar to that of \citet{Telfer2002}.  
In Fig.~\ref{fig:modelsed} we show the SED of \citet{Lusso2015} and the
continuum slopes of \citet{Telfer2002} and \citet{Scott2004}.

We calculated $t_{\mathrm{LyC}}$ for our sample AGNs with the same
procedure that we used in Section~\ref{subsec:lyctrans} but now changing
the intrinsic SED from the \citet{Lusso2015} AGN spectrum to one of the
two other SEDs described above. The difference between
$t_{\mathrm{LyC}}$ obtained assuming the \citet{Lusso2015} SED and that
with a different SED varies for each object.
For the case with the continuum slope of \citet{Telfer2002}
$t_{\mathrm{LyC}}$ values become higher, up to 0.1, and with a median of
0.026.  This happens because for the \citet{Telfer2002} SED the
intrinsic LyC  luminosity relative to the luminosity at 1450\AA\ is
smaller than for the \citet{Lusso2015} SED. On the other hand, if we use
the continuum slope from \citet{Scott2004} then the relative LyC
luminosity of the intrinsic SED is higher than that for the
\citet{Lusso2015} SED, and consequently the $t_{\mathrm{LyC}}$ values
become smaller, up to 0.17, with a median difference of 0.033. 

In these tests with the UV slopes by \citet{Telfer2002} and \citet{Scott2004}
we use continuum slopes as intrinsic SEDs, without
including emission lines. To check for the effect of emission lines we
also run a test using the quasar composite spectrum by
\citet{VandenBerk2001} which is based on spectra of SDSS quasars. This
SED is also shown in Fig.~\ref{fig:modelsed}.
Because this composite spectrum is not corrected for IGM absorption, at
$\lambda < 1200$\AA\, we use the continuum slope of \cite{Telfer2002},
namely $\alpha = -1.76$. In this case, $t_{\mathrm{LyC}}$ values become
larger than those we obtained with the \cite{Lusso2015} SED by a median
value of 0.046, with a maximum difference of 0.23.

These tests indicate that by altering the assumed intrinsic SED, the
estimate of $t_{\mathrm{LyC}}$ will change, and in most cases the
variation is less than 0.1 though there are some cases where the estimate 
varies by $\sim$0.2.
\citet{Lusso2015} provides uncertainties of their 
stacked spectrum estimated through bootstrap, and they are shown in 
Fig.~\ref{fig:modelsed}. As seen in the figure, the variations of SEDs 
in wavelengths shorter than the Lyman limit by considering UV slopes by 
\citet{Telfer2002} and \citet{Scott2004} are larger than the uncertainties 
in \citet{Lusso2015} spectrum. The possible changes in $t_{\mathrm{LyC}}$ 
due to uncertainties of \citet{Lusso2015} spectrum will be smaller than 
the variations considered above.
The intrinsic SED could possibly be different from AGN to AGN, and in
order to precisely determine the intrinsic SED at 
rest-frame $\lambda \lesssim 1200$\AA\ a high dispersion spectrum is
required to identify absorption by intervening H\textsc{i} clouds which is
not available for our sample AGNs.

When we calculate average LyC escape fractions $f_{\mathrm{esc}}$ using
stacked images, we use the intrinsic flux ratio between the LyC and UV
(Equation~\ref{eqn_fesc}) spectral regions. 
The intrinsic flux ratios for 
the continuum slopes from \cite{Scott2004} and \cite{Telfer2002} are 
0.738 and 0.455, respectively, 
while it is 0.601 for the \citet{Lusso2015} SED. 
Therefore, the mean $f_{\mathrm{esc}}$ value for our 38 AGNs at
$3.3<z<3.6$ will be 0.247 and 0.400, respectively, if we adopt the
slopes by \citet{Scott2004} and \citet{Telfer2002}. 

Recently \citet{VandenBerk2020} claimed that, based on \textit{GALEX}
photometry of SDSS quasars, EUV ($\lambda \lesssim 1000$\AA) slope of
these quasars are $-2.90\pm0.04$, much redder than those reported by
\citet{Telfer2002}, \citet{Scott2004}, and \citet{Lusso2015}. If the
intrinsic relative flux density of LyC to non-ionizing UV photon is
smaller than the fiducial model considered in this study, LyC escape
fraction could be higher, although LyC emissivity of the sample AGNs
(which is discussed in Section.~\ref{subsec:lycemissivity} will not
change.

\begin{figure}
 \includegraphics[width=\columnwidth]{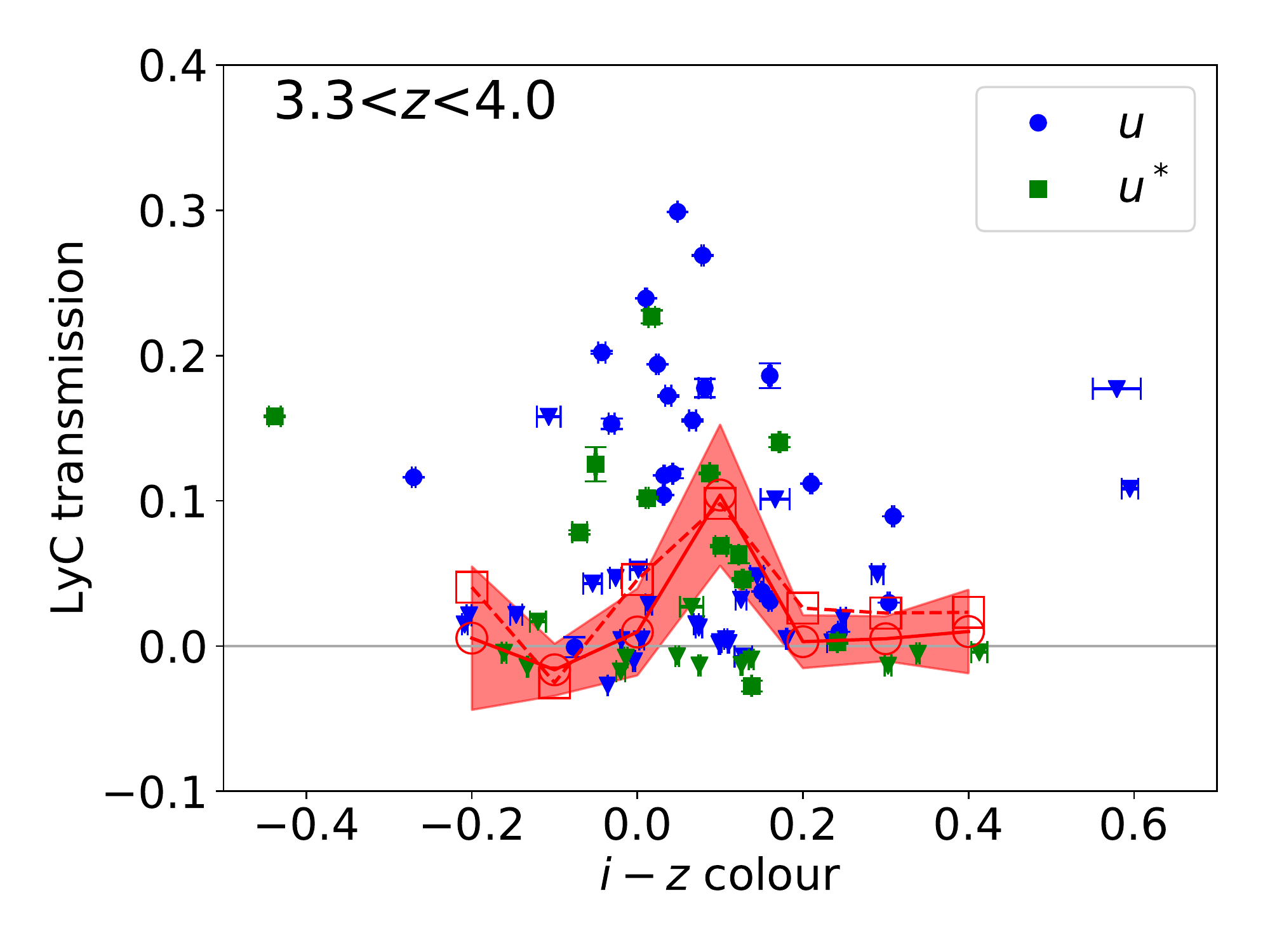}
 \caption{Distribution of $i-z$ colour and LyC transmission 
 for the nuclear-dominated sample AGNs at $3.4<z<4.0$. 
 The open circles and open squares indicate median and mean values in
 0.05 $i-z$ colour bins, respectively, and the shaded area represents
 the variation of the median values estimated by bootstrap resampling.
}
 \label{fig:izcolour}
\end{figure}

We also examine whether there is a trend that AGNs with redder
rest-frame UV slope, which might be caused by attenuation by dust, show
smaller LyC transmission compared to those with bluer UV slope. We use
$i-z$ colour as an indicator of rest-frame UV slope of the sample
AGNs\footnote{Here the effect of emission lines on
observed broad-band flux densities is ignored. \textit{i}-band contains
C\textsc{iv} emission line when the object is between 
$3.4 \lesssim z < 4.0$ (see Fig.~\ref{fig:filter_trans}), and the
variance of its strength may affect the observed $i-z$ colour.}, 
and do not find a significant correlation between the observed $i-z$
colours and the estimated LyC transmission values.
In Fig.~\ref{fig:izcolour} we show the distribution of LyC transmission
against the observed $i-z$ colours. We see that there are fewer objects
with relatively large LyC transmission among those with larger $i-z$
colours, and it may be caused by the presence of dust
attenuation. However, median and mean values do not indicate any
statistically significant correlation between the $i-z$ colours and the
LyC transmission values.

\subsection{Difference between broad-line and narrow-line AGNs}
\label{subsec:BLA_NLA}

The distribution of material in the vicinity of the AGN may play a role
in helping or hindering the escape of ionizing radiation.  Because BLAs
and NLAs are thought to represent different AGN viewing angles,
differences between these two types of AGN can help us determine if
geometry plays a role in the escape of ionizing radiation. 

The source of ionizing radiation from an AGN could be the accretion disk
surrounding its supermassive black hole, and if NLAs are those AGNs in
which a dusty torus obscures the broad-line region from our direct
observation, then ionizing radiation from NLAs could be heavily
attenuated by the intervening material.  Consequently, we might  see a
difference in average $f_\mathrm{esc}$ (and hence the distribution of
$t_{\mathrm{LyC}}$ values for individual sources) between BLAs and NLAs.
With this goal in mind, we split our AGN sample and search for a
difference in measured LyC transmission between the broad-line AGN (BLA)
subsample and narrow-line AGN (NLA) subsample. As described in
Section~\ref{subsec:extendedness}, we select AGNs which are classified
as `nuclear-dominated' by examining a difference between PSF magnitude
and magnitude with galaxy model (CModel) fitting to estimate LyC
transmission and to carry out stacking analysis, and all but two of 12
AGNs classified as NLA in literature are excluded. Here we use all of the
94 sample AGNs which include 20 AGNs classified as extended and all of
the 12 NLAs, and carried out the analysis to calculate
$t_{\mathrm{LyC}}$ in the same manner as described in
Section~\ref{subsec:lyctrans}, assuming the intrinsic SED to be the
fiducial AGN SED by \citet{Lusso2015}. For $U$-band photometry we use
PSF magnitudes for both nuclear-dominated and extended AGNs, to examine
LyC transmission from their central AGN which is expected to be a
point-like source.
In contrast to these results, \citet{Micheva2017a} examined LyC emission
from 14 AGNs in the SSA22 field in which half are type \textsc{ii} 
AGNs, and found that none of the 7 type \textsc{ii} AGNs has detectable
LyC emission, while LyC from two type \textsc{i} AGNs are detected.

\begin{figure*}
 \includegraphics[width=\columnwidth]{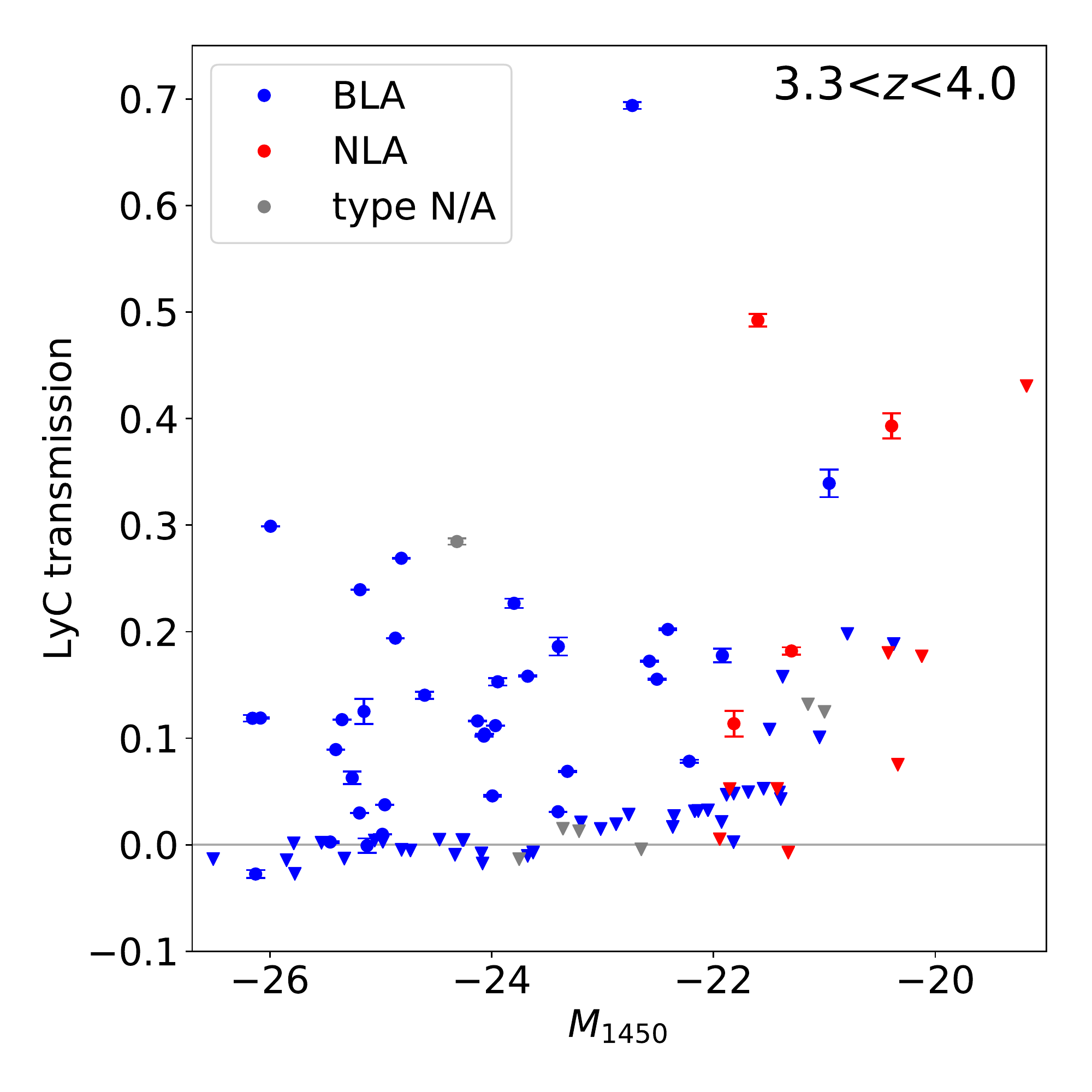}
 \includegraphics[width=\columnwidth]{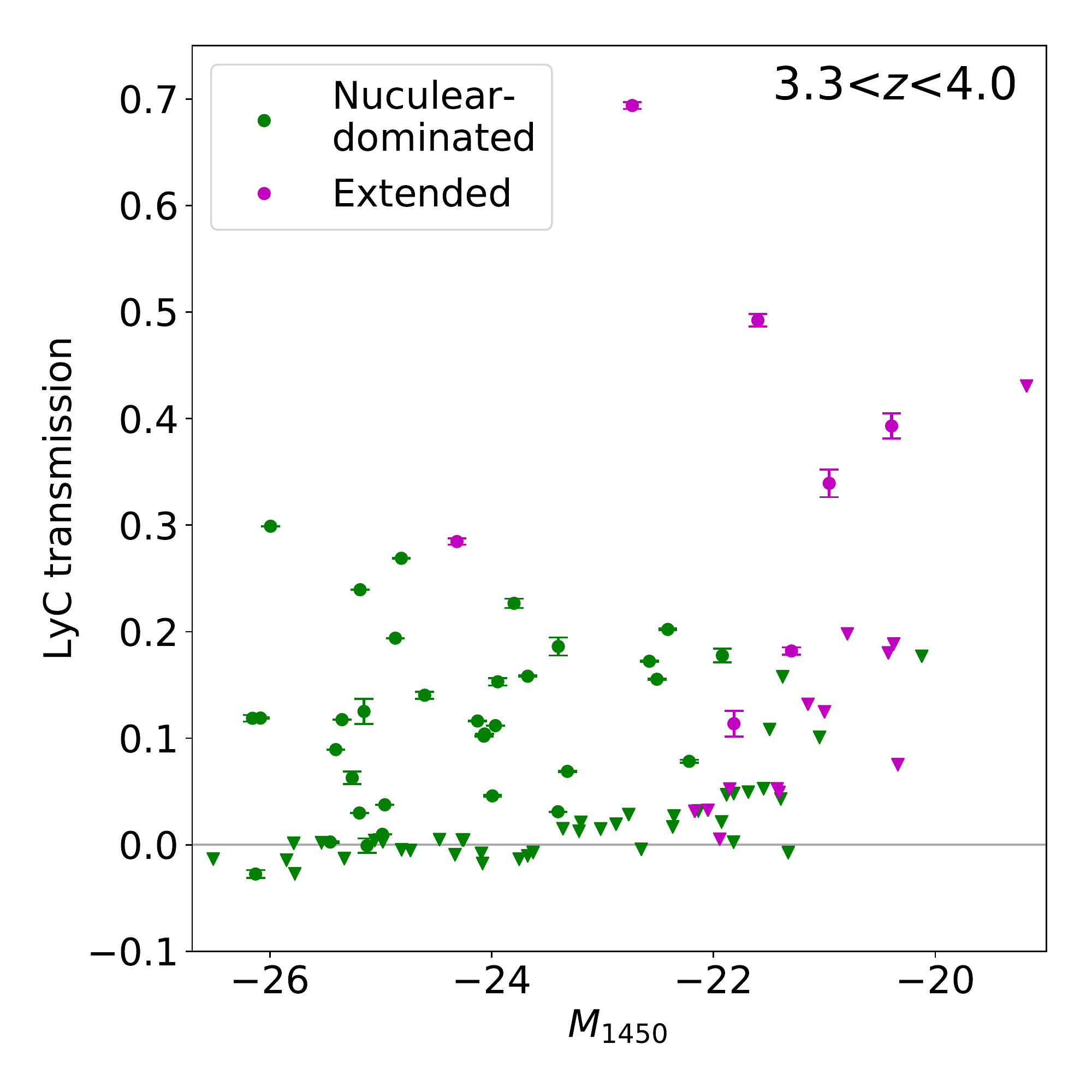}
 \caption{(left) LyC transmission $t_{\mathrm{LyC}}$ for the sample AGNs at
 $3.3<z<4.0$ plotted  against their absolute magnitudes at rest-frame
 1450\AA. Solid circles  are the measured $t_{\mathrm{LyC}}$, and
 downward-pointing triangles are 3$\sigma$ upper limits for the AGNs
 without detections in \fu or \fuS. BLAs and NLAs are shown with blue
 and red symbols, respectively, while AGNs without type information in
 the literature are shown with grey symbols.
(right) Same as the left figure, but the sample is divided by whether
 the UV flux from the object is dominated by photons from its nucleus or
 significant photons come from its host galaxy.}
 \label{fig:tlyc_bla_nla}
\end{figure*}

Fig.~\ref{fig:tlyc_bla_nla} shows $t_{\mathrm{LyC}}$ and absolute UV
magnitudes for the 94 sample AGNs, with classifications as BLA (blue),
NLA (red), and those without type information in the literature
(gray). Unfortunately, because the number of NLAs in our sample is
small, we cannot reach a solid conclusion at present. However, we see
that some of the highest $t_{\mathrm{LyC}}$ values are for NLAs, which
is at odds with the naive idea that NLAs may be AGNs with small LyC
escape fractions due to circum-AGN obscuration. Such trend is also seen
when we divide the sample into `nuclear-dominated' and `extended'; the
AGNs with extended morphology which are detected in \fu or \fuS filters
appear to show relatively high LyC transmission.

\begin{figure}
 \includegraphics[width=\columnwidth]{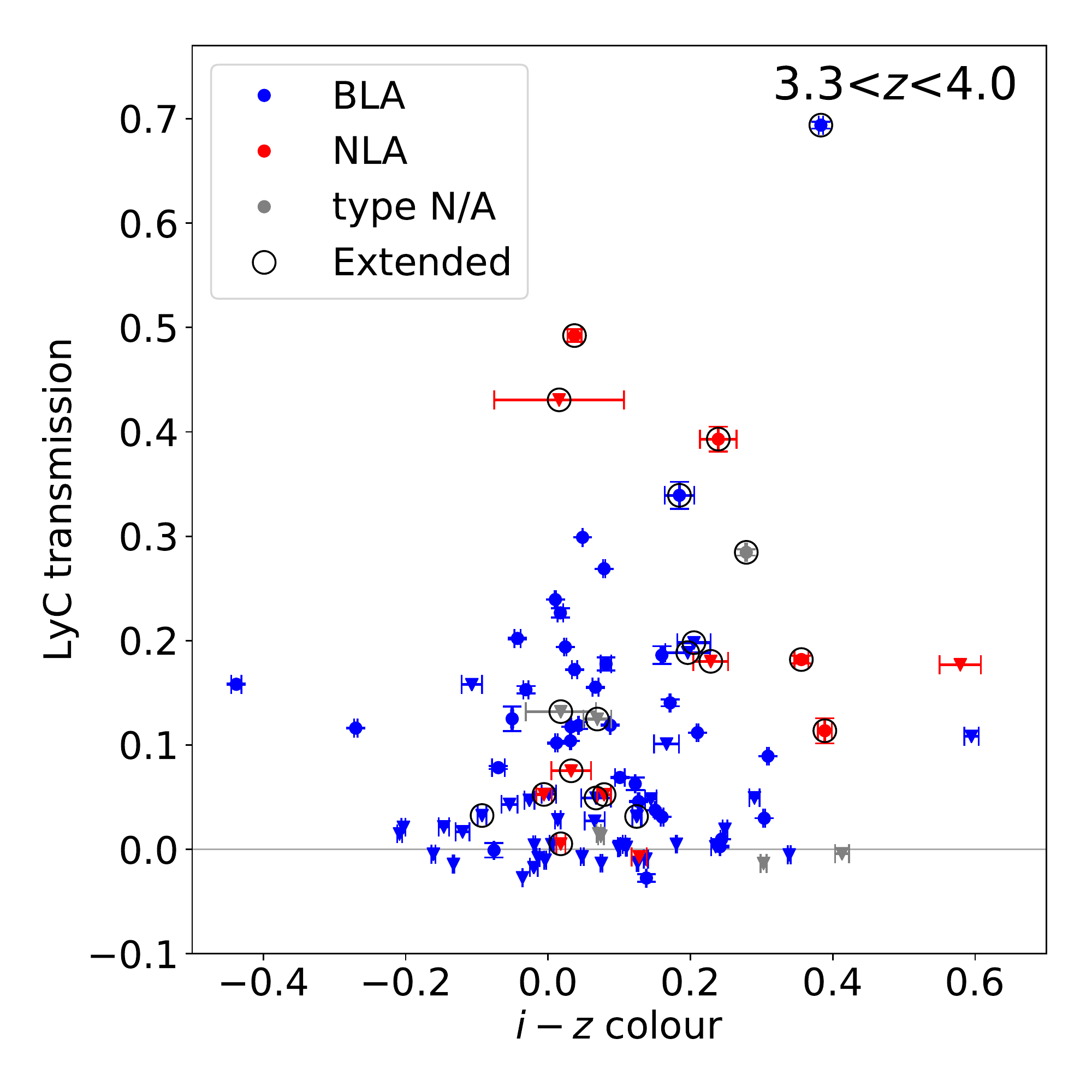}
 \caption{Distribution of $i-z$ colour and LyC transmission 
 for the sample AGNs at $3.4<z<4.0$. 
 Blue circles and red symbols represent BLAs and NLAs, respectively, 
 and open black circles indicate AGNs with extended morphology in
 \textit{i}-band.
}
 \label{fig:izcolour_BLANLA}
\end{figure}

Fig.~\ref{fig:izcolour_BLANLA} shows the distribution of 
$t_{\mathrm{LyC}}$ and $i-z$ colour for the sample AGNs, 
including NLAs and those with extended morphology. 
We see that many of the AGNs with large LyC transmission 
and red $i-z$ colour are extended, 
and some of them are classified as NLAs.
It would be reasonable to consider that stellar populations of the host
galaxies contribute significantly to UV flux densities in extended AGNs,
and because LyC escape fraction of star-forming galaxies at $z\sim 3$
selected by their rest-frame UV colours is generally believed to be
small \cite[less than 10\%; e.g.,][]{Steidel2018, Iwata2019}, it is
difficult to understand why extended AGNs show higher LyC transmission.
To make further progress on this issue larger AGN samples, particularly
those of NLAs, are needed. 

\subsection{Comparison with previous studies}

\citet{Cowie2009} used GALEX FUV measurements to search for LyC
from X-ray selected AGNs in a 0.9 deg$^2$ field. They found that only
small fraction of X-ray selected AGNs in the redshift range between 0.9
and 1.4 have detectable ionizing flux. They argue that the presence of
absorbers along the line of sight will extinguish the ionizing
radiation.
\citet{Cowie2009} also claim that only broad-line AGNs among their 32 
sample AGNs have detectable ionizing flux, which is at odds with our
finding (see Section~\ref{subsec:BLA_NLA}). 
They used an ionizing to non-ionizing flux ratio to estimate the
contribution to the ionizing background radiation and concluded that
AGNs provide insufficient ionizing radiation at $z > 3$.

\citet{Micheva2017a} used 14 AGNs in a redshift range $3.06<z<4.0$ in
the SSA22 field to search for LyC radiation with Subaru Suprime-Cam
narrow-band imaging. Their UV absolute magnitude ranges from $-25$ to
$-19$ and many of them are faint AGNs. Among the four LyC candidates,
two show offsets from the peak positions of non-ionizing UV, and 
their LyC radiation could come from stellar sources or they may be
contaminated by foreground sources. The \fesc\ of the remaining two
sources are estimated to be 0.31 and 0.73 when the median IGM
attenuation of their redshift is assumed.

\citet{Grazian2018} used deep optical spectra of 16 AGNs at
$3.6<z<4.2$. The $M_{\mathrm{1450}}$ of their sample AGNs ranges from
$-25.14$ to $-23.26$ which is fainter than typical SDSS quasars but 
brighter than UV luminosity ranges of the sample AGNs 
in \citet{Micheva2017a} and the present study. 
They detect LyC from all of their sample AGNs, and
derived the mean $f_{\mathrm{esc}} = 0.74$. They found no significant
difference between \fesc\ of their sample AGNs and the values for bright
quasars.
The method they use to
determine $f_{\mathrm{esc}}$ is different from the one used in
\citet{Micheva2017a} and this study; they used a flux ratio 
$f_\nu(\mathrm{900})$/$f_\nu(\mathrm{930})$ as $f_{\mathrm{esc}}$ and
used mean flux between rest-frame 892 and 905\AA\ and that for 
915 and 945\AA, respectively, for $f_\nu(\mathrm{900})$ and 
$f_\nu(\mathrm{930})$. The \fesc\ values \citet{Grazian2018} reported
tend to be higher than the values we present in this study, and the
difference may come from the different way to estimate \fesc.

The study by \citet{Romano2019} is based on a large sample of QSO
spectra obtained by SDSS DR14. They used 2508 QSOs at 
$3.6 \leq z \leq 4.6$ with UV absolute magnitude range 
$-29.0 \lesssim M_{\mathrm{1450}} \lesssim -26.0$, which include objects
at higher redshifts and with brighter UV luminosity range than those
used in this study. Their method to measure $f_{\mathrm{esc}}$ is the
same as \citet{Grazian2018}. They found $f_{\mathrm{esc}} =
0.49\pm0.36$, which they concluded to be consistent with the result by
\citet{Grazian2018}. 

\citet{Smith2020} used HST WFC3/UVIS imaging of the GOODS North and
South and the ERS fields to examine LyC emission from galaxies and AGNs
with spectroscopic redshifts at $2.26 < z< 4.3$. 
There are 17 AGNs in their sample. The rest-frame UV
absolute magnitudes of most of them is $M_{\mathrm{1500}}>-22$, fainter
than the sample AGNs in this study. Only one among them is detected with
the best estimate $f_{\mathrm{esc}} \simeq 28$--30\%.

Our results confirm the fact reported in these previous studies that
the average \fesc\ of high redshift AGNs is 
considerably smaller than unity. 
Although we find no clear luminosity dependence in
LyC transmission or LyC escape fraction, we would need a larger sample
of faint AGNs in order to further study the existence of luminosity
dependence. Also, we need sample of AGNs at different redshift ranges
with comparable size to investigate redshift evolution.

\subsection{Implications for contribution by AGNs to cosmic reionization}
\label{subsec:lycemissivity}

We can calculate the contribution of AGNs to
the volume-averaged ionizing radiation emissivity, by integrating the
LyC luminosity over a range of UV luminosities:

\begin{equation}
\epsilon_{\mathrm{LyC}} = \int
 (f_{\mathrm{LyC}}/f_{\mathrm{UV}})^{\mathrm{out}} L_{\mathrm{UV}} \phi_{\mathrm{UV}} d L_{\mathrm{UV}},
\label{eq_emissivity}
\end{equation}
where $L_{\mathrm{UV}}$ is the non-ionizing UV luminosity and
$\phi_{\mathrm{UV}}$ is the number density of AGNs in
the luminosity range (luminosity function). 
We use the UVLF (double power-law fit) for
the quasar sample at $3.6 \lesssim z \lesssim 4.2$ based on the HSC SSP
wide-layer imaging data by \citet{Akiyama2018}, 
with an assumption that there is no evolution in UVLF in 
the redshift range of the sample AGNs.
The mean redshift of the 38 nuclear-dominated sample AGNs at 
$3.3 < z < 3.6$ used to 
calculate $(f_{\mathrm{LyC}}/f_{\mathrm{UV}})^{\mathrm{out}}$ and
$f_{\mathrm{esc}}$ through the stacking analysis in
Section~\ref{sec:stacking} is 3.44, and we use the value to calculate
luminosity in equation~\ref{eq_emissivity}.
In Table~\ref{tab:emissivity} we show the calculated
$\epsilon_{\mathrm{LyC}}$ with three integration ranges. One is
$-26 < M_{\mathrm{1450}} < -20$ which is roughly the UV absolute
magnitude range of the sample AGNs, and the second one is
$-27 < M_{\mathrm{1450}} < -18$, an extended range, and the other is 
to integrate over the entire luminosity range.
Here we use $(f_{\mathrm{LyC}}/f_{\mathrm{UV}})^{\mathrm{out}}$ values
in Table~\ref{tab:stack} depending on $M_{\mathrm{1450}}$. 
If we use a
single $(f_{\mathrm{LyC}}/f_{\mathrm{UV}})^{\mathrm{out}}$ value of
0.182 independent of absolute UV magnitude, $\epsilon_{\mathrm{LyC}}$
decreases about $\sim$5\%  from the values in
Table~\ref{tab:emissivity}.

\begin{table*}
 \centering
\caption{LyC emissivity of $z\sim 3.5$ AGNs 
 using our measured LyC
 escape fraction 0.303$\pm$0.072 with three different UVLFs, compared
 with the ionizing background radiation at $z\sim 3.2$ based on QSO
 Ly$\alpha$ forest observations reported by \citet{Becker2013}.}
\label{tab:emissivity}
\begin{tabular}{ccccc}
\hline
 Integration range  & \multicolumn{4}{c}{$\epsilon_{\mathrm{LyC}}$ (10$^{24}$ erg/s/Hz/Mpc$^3$)} \\
\cline{2-5}
 & \citet{Akiyama2018} & \citet{Masters2012} & \citet{Giallongo2015} & \citet{Becker2013}$^a$\\
\hline
 $-26 < M_{\mathrm{1450}} < -20$ & $0.450\pm0.160$ & $9.19$ & $1.72$ & \\
 $-27 < M_{\mathrm{1450}} < -18$ & $0.558\pm0.203$ & $1.16$ & $2.01$ & \\
 $-\infty < M_{\mathrm{1450}} <\infty$ & $0.620\pm0.230$ & $1.43$ & $2.21$ & $8.15^{+13.38}_{-5.29}$ \\
\hline
\end{tabular}
\begin{flushleft}
\textit{Notes.}
$^a$: LyC emissivity based on Ly$\alpha$ forest observations.
\end{flushleft}
\end{table*}

\citet{Becker2013} reported $\epsilon_{\mathrm{LyC}}$ from all
sources based on QSO Ly$\alpha$ forest observations over $2<z<5$. The 
nominal value at $z = 3.2$ is 
$\epsilon_{\mathrm{LyC}} = 8.15^{+13.38}_{-5.29}\times10^{24}$
erg/s/Hz/Mpc$^3$.
Their error estimates include both statistical and systematic errors
which could arise during the course of the inference.
Our estimate of the contribution from AGNs to
the LyC emissivity based on the UVLF by \citet{Akiyama2018} is
5.5--7.6\% of the nominal value, depending on the integration range of
the UVLF.
Our results suggest that if LyC escape fraction of faint AGNs beyond the
luminosity range we study here remains the same level as those we
examine, the AGNs are the minor contributor to the background ionizing
radiation in the Universe at $z\sim 3.5$. 

The estimate of contribution in LyC emissivity by faint AGNs largely
depends on the assumed UVLF, especially on its faint-end slope.
The estimate of contribution in LyC emissivity by faint AGNs depends on
the assumed faint-end slope of the UVLF, and the
reported faint-end slopes of $z=3$--4 AGNs in the literature are largely
different (see Fig.~21 of \citet{Akiyama2018}). 
In Table~\ref{tab:emissivity}, in addition to the emissivity values
using the UVLF by \citet{Akiyama2018}, we also list values using the
UVLF by \citet{Masters2012} (for $z\sim3.2$ quasars) and that by
\citet{Giallongo2015} (for AGNs at $z=4$--4.5), assuming no evolution of
the UVLF. Because these UVLFs predict higher number density of faint
AGNs, the emissivity values using these UVLFs are 2 to 5 times higher
than the values with the UVLF by \citet{Akiyama2018}.
Also, because the redshift interval of the results by
\citet{Giallongo2015} we use in this study is $z=4$ to 4.5, if we adopt
a number density evolution $\phi(z) \propto 10^{-0.38 z}$ 
suggested by \citet{Schindler2019}, the number density at $z=3.4$ will
be a factor of $\sim$2 higher. In such case, the LyC emissivity by AGNs
could be as high as $\sim$40--50\% of the nominal value at $z=3.2$ by 
\citet{Becker2013}. 

\citet{Giallongo2015} argued that, based on their selection of faint AGN
candidates in the GOODS-S field with photometric redshift and X-ray
detection, the number density of faint AGNs at
$4 < z < 6.5$ is much higher than those estimated from the existing
luminosity function reported in literature, and suggested that LyC
emissivity by AGNs could provide ionizing photons sufficient to keep the
IGM ionized. For the redshift range $4 < z < 4.5$ their estimate of
$\epsilon_{\mathrm{LyC}}$ is $1.15 \times 10^{25}$ erg/s/Hz/Mpc$^3$,
which exceeds the nominal value of $\epsilon_{\mathrm{LyC}}$ from all
sources at $z\sim 4$ reported by \citet{Becker2013} ($9.62\times10^{24}$
erg/s/Hz/Mpc$^3$; 
see also \citet{Boutsia2021} for the updated UVLF and estimates on the
contributions to the LyC emissivity by faint AGNs).
We should note that recent results of quasar UVLF at $z\sim 4$ and higher
based on HSC SSP \citep{Akiyama2018, Matsuoka2018} suggest fewer number
density of faint AGNs. 
Also, our study suggest \fesc\ of high-z
AGNs could be 0.2--0.4 (Table~\ref{tab:stack}), much smaller than
unity which \citet{Giallongo2015} assumed.
If \fesc\ of AGNs at $z \gtrsim 4$ continues to be
less than unity as found in this study for $3.3 < z < 4$ AGNs, it
further reduces the contribution to the volume-averaged ionizing
radiation by AGNs at higher redshift.

\citet{Akiyama2018} argued that the difference between the UVLFs of
rest-frame UV selected AGNs and those of X-ray selected AGNs may come
from the fact that the latters are dominated by obscured AGNs in the
fainter part which are missed in the UV-selected sample
selection. Because LyC emission examined in this study is that from
nuclei of the sample AGNs, it would be reasonable to use the UVLF of
UV-selected AGNs to estimate LyC emissivity of the nuclear activities in
the AGNs at the epoch. However, we should be cautious that, since we do
not find a trend that LyC transmission estimates of NLA or AGNs with
extended UV emission are smaller than BLA or nuclear-dominated AGNs (see
Fig.~\ref{fig:tlyc_bla_nla}), contributions to LyC emissivity from
obscured AGNs or AGNs whose UV emission is dominated by those from host
galaxies could be non-negligible.

\section{Conclusions}

In this paper we assembled a sample of AGNs with spectroscopic redshift
between 3.3 and 4.0 in the four independent fields of the Deep layer of
the HSC SSP to constrain their LyC escape fraction (\fesc). 
Among the 94 AGNs we select 74 nuclear-dominated AGNs based on
\textit{i}-band photometry. We use deep $U$-band data from CLAUDS to
directly measure observed LyC flux densities. 
Our findings can be summarized as follows:

\begin{itemize}
 \item With a sample of 74 nuclear-dominated AGNs spanning in the UV absolute
       magnitude range $-26 \lesssim M_{\mathrm{1450}} \lesssim -20$,
       we do not find significant trend in \fesc\ with their UV
       luminosities. 
 \item By stacking 38 AGNs in the redshift range $3.3<z<3.6$ and
       assuming mean IGM attenuation and intrinsic LyC/UV flux ratio, we
       find the mean \fesc\ value of $0.303\pm0.072$. 
       The mean \fesc values for luminous AGNs 
       ($M_{\mathrm 1450} < -24$) and faint AGNs 
       ($M_{\mathrm 1450} > -24$) are $0.358\pm0.104$ and
       $0.253\pm0.101$, respectively. The difference between the value
       for luminous AGNs and that for faint AGNs is not significant.
 \item When we assume the UVLF by \citet{Akiyama2018}
       and use the average \fesc\ values, 
       the contribution to the LyC emissivity by AGNs at $z\sim 3.5$ is
       $\simeq$5--8\%. The estimate of contribution to the LyC emissivity
       by AGNs largely depends on the assumed UVLF, but if \fesc\
       remains much less than unity, by adopting the UVLF with flat
       faint-end slope recently reported 
       \citep{Akiyama2018, Matsuoka2018}, it is suggested that LyC
       emission from nuclei of AGNs at higher redshift would be a minor
       contributor to the volume-averaged ionizing radiation (i.e., not
       a major player of the cosmic reionization).
 \item We do not see a geometric effect on \fesc\ (no obvious BLA vs NLA
       difference), though a large NLA sample is needed to be definitive.
\end{itemize}

\section*{Acknowledgements}

II wishes to thank Dr. Konstantina Boutsia for providing several AGN
spectra in the COSMOS field used in this study, Dr. Wanqiu He for
additional redshift information of AGNs in the XMM-LSS field, and
Dr. Masayuki Tanaka for valuable suggestions.
We are grateful to the anonymous referee for valuable comments.

This work was supported by JSPS KAKENHI Grant Numbers 18740114,
24244018, and 17H01114 and by a Discovery Grant from the Natural
Sciences and Engineering Research Council of Canada. 

The Hyper Suprime-Cam (HSC) collaboration includes the astronomical communities of Japan and Taiwan, and Princeton University. The HSC instrumentation and software were developed by NAOJ, the Kavli Institute for the Physics and Mathematics of the Universe (Kavli IPMU), the University of Tokyo, the High Energy Accelerator Research Organization (KEK), the Academia Sinica Institute for Astronomy and Astrophysics in Taiwan (ASIAA), and Princeton University. Funding was contributed by the FIRST program from Japanese Cabinet Office, the Ministry of Education, Culture, Sports, Science and Technology (MEXT), the Japan Society for the Promotion of Science (JSPS), Japan Science and Technology Agency (JST), the Toray Science Foundation, NAOJ, Kavli IPMU, KEK, ASIAA, and Princeton University.

This work is based on observations obtained with MegaPrime/MegaCam, a joint project of CFHT and CEA/DAPNIA, at the Canada-France-Hawaii Telescope (CFHT) which is operated by the National Research Council (NRC) of Canada, the Institut National des Science de l'Univers of the Centre National de la Recherche Scientifique (CNRS) of France, and the University of Hawaii. This research uses data obtained through the Telescope Access Program (TAP), which has been funded by the National Astronomical Observatories, Chinese Academy of Sciences, and the Special Fund for Astronomy from the Ministry of Finance. This work uses data products from TERAPIX and the Canadian Astronomy Data Centre. It was carried out using resources from Compute Canada and Canadian Advanced Network For Astrophysical Research (CANFAR) infrastructure.

This paper makes use of software developed for the Large Synoptic Survey Telescope. We thank the LSST Project for making their code available as free software at http://dm.lsstcorp.org.

The Pan-STARRS1 Surveys (PS1) have been made possible through contributions of the Institute for Astronomy, the University of Hawaii, the Pan-STARRS Project Office, the Max-Planck Society and its participating institutes, the Max Planck Institute for Astronomy, Heidelberg and the Max Planck Institute for Extraterrestrial Physics, Garching, The Johns Hopkins University, Durham University, the University of Edinburgh, Queen's University Belfast, the Harvard-Smithsonian Center for Astrophysics, the Las Cumbres Observatory Global Telescope Network Incorporated, the National Central University of Taiwan, the Space Telescope Science Institute, the National Aeronautics and Space Administration under Grant No. NNX08AR22G issued through the Planetary Science Division of the NASA Science Mission Directorate, the National Science Foundation under Grant No. AST-1238877, the University of Maryland, and Eotvos Lorand University (ELTE).

This research partly uses database of Sloan Digital Sky Survey
(SDSS). Funding for the SDSS and SDSS-II has been provided by the Alfred
P. Sloan Foundation, the Participating Institutions, the National
Science Foundation, the U.S. Department of Energy, the National
Aeronautics and Space Administration, the Japanese Monbukagakusho, the
Max Planck Society, and the Higher Education Funding Council for
England. The SDSS Web Site is http://www.sdss.org/.


This research made use of Astropy, a community-developed core Python
package for Astronomy.


We wish to express our gratitude to the indigenous Hawaiian community
for their understanding of the significant role of the summit of
Maunakea in astronomical research.


\section*{Data Availability}

Hyper Suprime-Cam Subaru Strategic Program Public Data are available at
https://hsc-release.mtk.nao.ac.jp/. The data underlying this article
will be shared on reasonable request to the corresponding author.



\bibliographystyle{mnras}
\bibliography{LyCAGN}

\begin{thebibliography}{}
\makeatletter
\relax
\def\mn@urlcharsother{\let\do\@makeother \do\$\do\&\do\#\do\^\do\_\do\%\do\~}
\def\mn@doi{\begingroup\mn@urlcharsother \@ifnextchar [ {\mn@doi@}
  {\mn@doi@[]}}
\def\mn@doi@[#1]#2{\def\@tempa{#1}\ifx\@tempa\@empty \href
  {http://dx.doi.org/#2} {doi:#2}\else \href {http://dx.doi.org/#2} {#1}\fi
  \endgroup}
\def\mn@eprint#1#2{\mn@eprint@#1:#2::\@nil}
\def\mn@eprint@arXiv#1{\href {http://arxiv.org/abs/#1} {{\tt arXiv:#1}}}
\def\mn@eprint@dblp#1{\href {http://dblp.uni-trier.de/rec/bibtex/#1.xml}
  {dblp:#1}}
\def\mn@eprint@#1:#2:#3:#4\@nil{\def\@tempa {#1}\def\@tempb {#2}\def\@tempc
  {#3}\ifx \@tempc \@empty \let \@tempc \@tempb \let \@tempb \@tempa \fi \ifx
  \@tempb \@empty \def\@tempb {arXiv}\fi \@ifundefined
  {mn@eprint@\@tempb}{\@tempb:\@tempc}{\expandafter \expandafter \csname
  mn@eprint@\@tempb\endcsname \expandafter{\@tempc}}}

\bibitem[\protect\citeauthoryear{{Aihara} et~al.,}{{Aihara}
  et~al.}{2018}]{Aihara2018}
{Aihara} H.,  et~al., 2018, \mn@doi [\pasj] {10.1093/pasj/psx066}, \href
  {http://ads.nao.ac.jp/abs/2018PASJ...70S...4A} {70, S4}

\bibitem[\protect\citeauthoryear{{Aihara} et~al.,}{{Aihara}
  et~al.}{2019}]{Aihara2019}
{Aihara} H.,  et~al., 2019, \mn@doi [\pasj] {10.1093/pasj/psz103}, \href
  {https://ui.adsabs.harvard.edu/abs/2019PASJ...71..114A} {71, 114}

\bibitem[\protect\citeauthoryear{{Akiyama} et~al.,}{{Akiyama}
  et~al.}{2015}]{Akiyama2015}
{Akiyama} M.,  et~al., 2015, \mn@doi [\pasj] {10.1093/pasj/psv050}, \href
  {http://adsabs.harvard.edu/abs/2015PASJ...67...82A} {67, 82}

\bibitem[\protect\citeauthoryear{{Akiyama} et~al.,}{{Akiyama}
  et~al.}{2018}]{Akiyama2018}
{Akiyama} M.,  et~al., 2018, \mn@doi [\pasj] {10.1093/pasj/psx091}, \href
  {https://ui.adsabs.harvard.edu/abs/2018PASJ...70S..34A} {70, S34}

\bibitem[\protect\citeauthoryear{{Becker} \& {Bolton}}{{Becker} \&
  {Bolton}}{2013}]{Becker2013}
{Becker} G.~D.,  {Bolton} J.~S.,  2013, \mn@doi [\mnras]
  {10.1093/mnras/stt1610}, \href
  {http://adsabs.harvard.edu/abs/2013MNRAS.436.1023B} {436, 1023}

\bibitem[\protect\citeauthoryear{{Bertin} \& {Arnouts}}{{Bertin} \&
  {Arnouts}}{1996}]{Bertin1996}
{Bertin} E.,  {Arnouts} S.,  1996, \mn@doi [\aaps] {10.1051/aas:1996164}, \href
  {http://adsabs.harvard.edu/abs/1996A%26AS..117..393B} {117, 393}

\bibitem[\protect\citeauthoryear{{Bosch} et~al.,}{{Bosch}
  et~al.}{2018}]{Bosch2018}
{Bosch} J.,  et~al., 2018, \mn@doi [\pasj] {10.1093/pasj/psx080}, \href
  {https://ui.adsabs.harvard.edu/abs/2018PASJ...70S...5B} {70, S5}

\bibitem[\protect\citeauthoryear{{Boutsia}, {Grazian}, {Giallongo}, {Fiore}  \&
  {Civano}}{{Boutsia} et~al.}{2018}]{Boutsia2018}
{Boutsia} K.,  {Grazian} A.,  {Giallongo} E.,  {Fiore} F.,   {Civano} F.,
  2018, \mn@doi [\apj] {10.3847/1538-4357/aae6c7}, \href
  {https://ui.adsabs.harvard.edu/abs/2018ApJ...869...20B} {869, 20}

\bibitem[\protect\citeauthoryear{{Boutsia} et~al.,}{{Boutsia}
  et~al.}{2021}]{Boutsia2021}
{Boutsia} K.,  et~al., 2021, arXiv e-prints, \href
  {https://ui.adsabs.harvard.edu/abs/2021arXiv210310446B} {p. arXiv:2103.10446}

\bibitem[\protect\citeauthoryear{Bradley et~al.,}{Bradley
  et~al.}{2019}]{larry_bradley_2019_2533376}
Bradley L.,  et~al., 2019, astropy/photutils: v0.6,
  \mn@doi{10.5281/zenodo.2533376}

\bibitem[\protect\citeauthoryear{{Brusa} et~al.,}{{Brusa}
  et~al.}{2010}]{Brusa2010}
{Brusa} M.,  et~al., 2010, \mn@doi [\apj] {10.1088/0004-637X/716/1/348}, \href
  {http://adsabs.harvard.edu/abs/2010ApJ...716..348B} {716, 348}

\bibitem[\protect\citeauthoryear{{Civano} et~al.,}{{Civano}
  et~al.}{2012}]{Civano2012}
{Civano} F.,  et~al., 2012, \mn@doi [\apjs] {10.1088/0067-0049/201/2/30}, \href
  {https://ui.adsabs.harvard.edu/abs/2012ApJS..201...30C} {201, 30}

\bibitem[\protect\citeauthoryear{{Coil} et~al.,}{{Coil}
  et~al.}{2011}]{Coil2011}
{Coil} A.~L.,  et~al., 2011, \mn@doi [\apj] {10.1088/0004-637X/741/1/8}, \href
  {http://adsabs.harvard.edu/abs/2011ApJ...741....8C} {741, 8}

\bibitem[\protect\citeauthoryear{{Cool} et~al.,}{{Cool}
  et~al.}{2013}]{Cool2013}
{Cool} R.~J.,  et~al., 2013, \mn@doi [\apj] {10.1088/0004-637X/767/2/118},
  \href {http://adsabs.harvard.edu/abs/2013ApJ...767..118C} {767, 118}

\bibitem[\protect\citeauthoryear{{Cowie}, {Barger}  \& {Trouille}}{{Cowie}
  et~al.}{2009}]{Cowie2009}
{Cowie} L.~L.,  {Barger} A.~J.,   {Trouille} L.,  2009, \mn@doi [\apj]
  {10.1088/0004-637X/692/2/1476}, \href
  {https://ui.adsabs.harvard.edu/abs/2009ApJ...692.1476C} {692, 1476}

\bibitem[\protect\citeauthoryear{{Cristiani}, {Serrano}, {Fontanot}, {Vanzella}
   \& {Monaco}}{{Cristiani} et~al.}{2016}]{Cristiani2016}
{Cristiani} S.,  {Serrano} L.~M.,  {Fontanot} F.,  {Vanzella} E.,   {Monaco}
  P.,  2016, \mn@doi [\mnras] {10.1093/mnras/stw1810}, \href
  {https://ui.adsabs.harvard.edu/abs/2016MNRAS.462.2478C} {462, 2478}

\bibitem[\protect\citeauthoryear{{Cucciati} et~al.,}{{Cucciati}
  et~al.}{2017}]{Cucciati2017}
{Cucciati} O.,  et~al., 2017, \mn@doi [\aap] {10.1051/0004-6361/201630113},
  \href {http://adsabs.harvard.edu/abs/2017A%26A...602A..15C} {602, A15}

\bibitem[\protect\citeauthoryear{{Dressler}, {Henry}, {Martin}, {Sawicki},
  {McCarthy}  \& {Villaneuva}}{{Dressler} et~al.}{2015}]{Dressler2015}
{Dressler} A.,  {Henry} A.,  {Martin} C.~L.,  {Sawicki} M.,  {McCarthy} P.,
  {Villaneuva} E.,  2015, \mn@doi [\apj] {10.1088/0004-637X/806/1/19}, \href
  {http://adsabs.harvard.edu/abs/2015ApJ...806...19D} {806, 19}

\bibitem[\protect\citeauthoryear{{Finkelstein} et~al.,}{{Finkelstein}
  et~al.}{2015}]{Finkelstein2015}
{Finkelstein} S.~L.,  et~al., 2015, \mn@doi [\apj]
  {10.1088/0004-637X/810/1/71}, \href
  {http://adsabs.harvard.edu/abs/2015ApJ...810...71F} {810, 71}

\bibitem[\protect\citeauthoryear{{Giallongo} et~al.,}{{Giallongo}
  et~al.}{2015}]{Giallongo2015}
{Giallongo} E.,  et~al., 2015, \mn@doi [\aap] {10.1051/0004-6361/201425334},
  \href {http://adsabs.harvard.edu/abs/2015A%26A...578A..83G} {578, A83}

\bibitem[\protect\citeauthoryear{{Giallongo} et~al.,}{{Giallongo}
  et~al.}{2019}]{Giallongo2019}
{Giallongo} E.,  et~al., 2019, \mn@doi [\apj] {10.3847/1538-4357/ab39e1}, \href
  {https://ui.adsabs.harvard.edu/abs/2019ApJ...884...19G} {884, 19}

\bibitem[\protect\citeauthoryear{{Grazian} et~al.,}{{Grazian}
  et~al.}{2018}]{Grazian2018}
{Grazian} A.,  et~al., 2018, \mn@doi [\aap] {10.1051/0004-6361/201732385},
  \href {http://adsabs.harvard.edu/abs/2018A%26A...613A..44G} {613, A44}

\bibitem[\protect\citeauthoryear{{Grazian} et~al.,}{{Grazian}
  et~al.}{2020}]{Grazian2020}
{Grazian} A.,  et~al., 2020, \mn@doi [\apj] {10.3847/1538-4357/ab99a3}, \href
  {https://ui.adsabs.harvard.edu/abs/2020ApJ...897...94G} {897, 94}

\bibitem[\protect\citeauthoryear{{Hasinger} et~al.,}{{Hasinger}
  et~al.}{2018}]{Hasinger2018}
{Hasinger} G.,  et~al., 2018, \mn@doi [\apj] {10.3847/1538-4357/aabacf}, \href
  {http://adsabs.harvard.edu/abs/2018ApJ...858...77H} {858, 77}

\bibitem[\protect\citeauthoryear{{Hennawi} \& {Prochaska}}{{Hennawi} \&
  {Prochaska}}{2007}]{Hennawi2007}
{Hennawi} J.~F.,  {Prochaska} J.~X.,  2007, \mn@doi [\apj] {10.1086/509770},
  \href {https://ui.adsabs.harvard.edu/abs/2007ApJ...655..735H} {655, 735}

\bibitem[\protect\citeauthoryear{{Inoue} \& {Iwata}}{{Inoue} \&
  {Iwata}}{2008}]{Inoue2008}
{Inoue} A.~K.,  {Iwata} I.,  2008, \mn@doi [\mnras]
  {10.1111/j.1365-2966.2008.13350.x}, \href
  {http://adsabs.harvard.edu/abs/2008MNRAS.387.1681I} {387, 1681}

\bibitem[\protect\citeauthoryear{{Inoue}, {Iwata}  \& {Deharveng}}{{Inoue}
  et~al.}{2006}]{Inoue2006}
{Inoue} A.~K.,  {Iwata} I.,   {Deharveng} J.-M.,  2006, \mn@doi [\mnras]
  {10.1111/j.1745-3933.2006.00195.x}, \href
  {https://ui.adsabs.harvard.edu/abs/2006MNRAS.371L...1I} {371, L1}

\bibitem[\protect\citeauthoryear{{Inoue}, {Shimizu}, {Iwata}  \&
  {Tanaka}}{{Inoue} et~al.}{2014}]{Inoue2014}
{Inoue} A.~K.,  {Shimizu} I.,  {Iwata} I.,   {Tanaka} M.,  2014, \mn@doi
  [\mnras] {10.1093/mnras/stu936}, \href
  {http://adsabs.harvard.edu/abs/2014MNRAS.442.1805I} {442, 1805}

\bibitem[\protect\citeauthoryear{{Ishigaki}, {Kawamata}, {Ouchi}, {Oguri},
  {Shimasaku}  \& {Ono}}{{Ishigaki} et~al.}{2015}]{Ishigaki2015}
{Ishigaki} M.,  {Kawamata} R.,  {Ouchi} M.,  {Oguri} M.,  {Shimasaku} K.,
  {Ono} Y.,  2015, \mn@doi [\apj] {10.1088/0004-637X/799/1/12}, \href
  {https://ui.adsabs.harvard.edu/abs/2015ApJ...799...12I} {799, 12}

\bibitem[\protect\citeauthoryear{{Iwata}, {Inoue}, {Micheva}, {Matsuda}  \&
  {Yamada}}{{Iwata} et~al.}{2019}]{Iwata2019}
{Iwata} I.,  {Inoue} A.~K.,  {Micheva} G.,  {Matsuda} Y.,   {Yamada} T.,  2019,
  arXiv e-prints, \href {https://ui.adsabs.harvard.edu/abs/2019arXiv190711113I}
  {p. arXiv:1907.11113}

\bibitem[\protect\citeauthoryear{{Kulkarni}, {Worseck}  \&
  {Hennawi}}{{Kulkarni} et~al.}{2019}]{Kulkarni2019}
{Kulkarni} G.,  {Worseck} G.,   {Hennawi} J.~F.,  2019, \mn@doi [\mnras]
  {10.1093/mnras/stz1493}, \href
  {https://ui.adsabs.harvard.edu/abs/2019MNRAS.488.1035K} {488, 1035}

\bibitem[\protect\citeauthoryear{{Le F{\`e}vre} et~al.,}{{Le F{\`e}vre}
  et~al.}{2013}]{LeFevre2013}
{Le F{\`e}vre} O.,  et~al., 2013, \mn@doi [\aap] {10.1051/0004-6361/201322179},
  \href {http://adsabs.harvard.edu/abs/2013A%26A...559A..14L} {559, A14}

\bibitem[\protect\citeauthoryear{{Lusso}, {Worseck}, {Hennawi}, {Prochaska},
  {Vignali}, {Stern}  \& {O'Meara}}{{Lusso} et~al.}{2015}]{Lusso2015}
{Lusso} E.,  {Worseck} G.,  {Hennawi} J.~F.,  {Prochaska} J.~X.,  {Vignali} C.,
   {Stern} J.,   {O'Meara} J.~M.,  2015, \mn@doi [\mnras]
  {10.1093/mnras/stv516}, \href
  {http://adsabs.harvard.edu/abs/2015MNRAS.449.4204L} {449, 4204}

\bibitem[\protect\citeauthoryear{{Marchesi} et~al.,}{{Marchesi}
  et~al.}{2016}]{Marchesi2016}
{Marchesi} S.,  et~al., 2016, \mn@doi [\apj] {10.3847/0004-637X/817/1/34},
  \href {http://adsabs.harvard.edu/abs/2016ApJ...817...34M} {817, 34}

\bibitem[\protect\citeauthoryear{{Masters} et~al.,}{{Masters}
  et~al.}{2012}]{Masters2012}
{Masters} D.,  et~al., 2012, \mn@doi [\apj] {10.1088/0004-637X/755/2/169},
  \href {http://adsabs.harvard.edu/abs/2012ApJ...755..169M} {755, 169}

\bibitem[\protect\citeauthoryear{{Matsuoka} et~al.,}{{Matsuoka}
  et~al.}{2018}]{Matsuoka2018}
{Matsuoka} Y.,  et~al., 2018, \mn@doi [\apj] {10.3847/1538-4357/aaee7a}, \href
  {https://ui.adsabs.harvard.edu/abs/2018ApJ...869..150M} {869, 150}

\bibitem[\protect\citeauthoryear{{Menzel} et~al.,}{{Menzel}
  et~al.}{2016}]{Menzel2016}
{Menzel} M.-L.,  et~al., 2016, \mn@doi [\mnras] {10.1093/mnras/stv2749}, \href
  {http://adsabs.harvard.edu/abs/2016MNRAS.457..110M} {457, 110}

\bibitem[\protect\citeauthoryear{{Micheva}, {Iwata}  \& {Inoue}}{{Micheva}
  et~al.}{2017a}]{Micheva2017a}
{Micheva} G.,  {Iwata} I.,   {Inoue} A.~K.,  2017a, \mn@doi [\mnras]
  {10.1093/mnras/stw1329}, \href
  {http://adsabs.harvard.edu/abs/2017MNRAS.465..302M} {465, 302}

\bibitem[\protect\citeauthoryear{{Micheva}, {Iwata}, {Inoue}, {Matsuda},
  {Yamada}  \& {Hayashino}}{{Micheva} et~al.}{2017b}]{Micheva2017b}
{Micheva} G.,  {Iwata} I.,  {Inoue} A.~K.,  {Matsuda} Y.,  {Yamada} T.,
  {Hayashino} T.,  2017b, \mn@doi [\mnras] {10.1093/mnras/stw2700}, \href
  {http://adsabs.harvard.edu/abs/2017MNRAS.465..316M} {465, 316}

\bibitem[\protect\citeauthoryear{{Nakajima}, {Ellis}, {Robertson}, {Tang}  \&
  {Stark}}{{Nakajima} et~al.}{2020}]{Nakajima2020}
{Nakajima} K.,  {Ellis} R.~S.,  {Robertson} B.~E.,  {Tang} M.,   {Stark} D.~P.,
   2020, \mn@doi [\apj] {10.3847/1538-4357/ab6604}, \href
  {https://ui.adsabs.harvard.edu/abs/2020ApJ...889..161N} {889, 161}

\bibitem[\protect\citeauthoryear{{Onoue} et~al.,}{{Onoue}
  et~al.}{2017}]{Onoue2017}
{Onoue} M.,  et~al., 2017, \mn@doi [\apjl] {10.3847/2041-8213/aa8cc6}, \href
  {https://ui.adsabs.harvard.edu/abs/2017ApJ...847L..15O} {847, L15}

\bibitem[\protect\citeauthoryear{{P{\^a}ris} et~al.,}{{P{\^a}ris}
  et~al.}{2018}]{Paris2018}
{P{\^a}ris} I.,  et~al., 2018, \mn@doi [\aap] {10.1051/0004-6361/201732445},
  \href {https://ui.adsabs.harvard.edu/\#abs/2018A&A...613A..51P} {613, A51}

\bibitem[\protect\citeauthoryear{{Pentericci}, {McLure}, {Franzetti}, {Garilli}
   \& {the VANDELS team}}{{Pentericci} et~al.}{2018}]{Pentericci2018}
{Pentericci} L.,  {McLure} R.~J.,  {Franzetti} P.,  {Garilli} B.,   {the
  VANDELS team} 2018, arXiv e-prints, \href
  {https://ui.adsabs.harvard.edu/\#abs/2018arXiv181105298P} {p.
  arXiv:1811.05298}

\bibitem[\protect\citeauthoryear{{Prochaska}, {Worseck}  \&
  {O'Meara}}{{Prochaska} et~al.}{2009}]{Prochaska2009}
{Prochaska} J.~X.,  {Worseck} G.,   {O'Meara} J.~M.,  2009, \mn@doi [\apjl]
  {10.1088/0004-637X/705/2/L113}, \href
  {https://ui.adsabs.harvard.edu/abs/2009ApJ...705L.113P} {705, L113}

\bibitem[\protect\citeauthoryear{{Prochaska} et~al.,}{{Prochaska}
  et~al.}{2013}]{Prochaska2013}
{Prochaska} J.~X.,  et~al., 2013, \mn@doi [\apj] {10.1088/0004-637X/776/2/136},
  \href {https://ui.adsabs.harvard.edu/abs/2013ApJ...776..136P} {776, 136}

\bibitem[\protect\citeauthoryear{{Robertson} et~al.,}{{Robertson}
  et~al.}{2013}]{Robertson2013}
{Robertson} B.~E.,  et~al., 2013, \mn@doi [\apj] {10.1088/0004-637X/768/1/71},
  \href {https://ui.adsabs.harvard.edu/abs/2013ApJ...768...71R} {768, 71}

\bibitem[\protect\citeauthoryear{{Romano}, {Grazian}, {Giallongo}, {Cristiani},
  {Fontanot}, {Boutsia}, {Fiore}  \& {Menci}}{{Romano}
  et~al.}{2019}]{Romano2019}
{Romano} M.,  {Grazian} A.,  {Giallongo} E.,  {Cristiani} S.,  {Fontanot} F.,
  {Boutsia} K.,  {Fiore} F.,   {Menci} N.,  2019, \mn@doi [\aap]
  {10.1051/0004-6361/201935550}, \href
  {https://ui.adsabs.harvard.edu/abs/2019A&A...632A..45R} {632, A45}

\bibitem[\protect\citeauthoryear{{Sawicki} \& {Thompson}}{{Sawicki} \&
  {Thompson}}{2006}]{Sawicki2006}
{Sawicki} M.,  {Thompson} D.,  2006, \mn@doi [\apj] {10.1086/500999}, \href
  {https://ui.adsabs.harvard.edu/abs/2006ApJ...642..653S} {642, 653}

\bibitem[\protect\citeauthoryear{Sawicki et~al.,}{Sawicki
  et~al.}{2019}]{Sawicki2019}
Sawicki M.,  et~al., 2019, \mn@doi [Monthly Notices of the Royal Astronomical
  Society] {10.1093/mnras/stz2522}, 489, 5202

\bibitem[\protect\citeauthoryear{{Schindler} et~al.,}{{Schindler}
  et~al.}{2019}]{Schindler2019}
{Schindler} J.-T.,  et~al., 2019, \mn@doi [\apj] {10.3847/1538-4357/aaf86c},
  \href {https://ui.adsabs.harvard.edu/abs/2019ApJ...871..258S} {871, 258}

\bibitem[\protect\citeauthoryear{{Schlafly} \& {Finkbeiner}}{{Schlafly} \&
  {Finkbeiner}}{2011}]{Schlafly2011}
{Schlafly} E.~F.,  {Finkbeiner} D.~P.,  2011, \mn@doi [\apj]
  {10.1088/0004-637X/737/2/103}, \href
  {http://adsabs.harvard.edu/abs/2011ApJ...737..103S} {737, 103}

\bibitem[\protect\citeauthoryear{{Schlegel}, {Finkbeiner}  \&
  {Davis}}{{Schlegel} et~al.}{1998}]{Schlegel1998}
{Schlegel} D.~J.,  {Finkbeiner} D.~P.,   {Davis} M.,  1998, \mn@doi [\apj]
  {10.1086/305772}, \href {http://adsabs.harvard.edu/abs/1998ApJ...500..525S}
  {500, 525}

\bibitem[\protect\citeauthoryear{{Scott}, {Bechtold}, {Dobrzycki}  \&
  {Kulkarni}}{{Scott} et~al.}{2000}]{Scott2000}
{Scott} J.,  {Bechtold} J.,  {Dobrzycki} A.,   {Kulkarni} V.~P.,  2000, \mn@doi
  [\apjs] {10.1086/317340}, \href
  {https://ui.adsabs.harvard.edu/abs/2000ApJS..130...67S} {130, 67}

\bibitem[\protect\citeauthoryear{{Scott}, {Kriss}, {Brotherton}, {Green},
  {Hutchings}, {Shull}  \& {Zheng}}{{Scott} et~al.}{2004}]{Scott2004}
{Scott} J.~E.,  {Kriss} G.~A.,  {Brotherton} M.,  {Green} R.~F.,  {Hutchings}
  J.,  {Shull} J.~M.,   {Zheng} W.,  2004, \mn@doi [\apj] {10.1086/422336},
  \href {https://ui.adsabs.harvard.edu/abs/2004ApJ...615..135S} {615, 135}

\bibitem[\protect\citeauthoryear{{Smith} et~al.,}{{Smith}
  et~al.}{2020}]{Smith2020}
{Smith} B.~M.,  et~al., 2020, \mn@doi [\apj] {10.3847/1538-4357/ab8811}, \href
  {https://ui.adsabs.harvard.edu/abs/2020ApJ...897...41S} {897, 41}

\bibitem[\protect\citeauthoryear{{Steidel}, {Bogosavljevi{\'c}}, {Shapley},
  {Reddy}, {Rudie}, {Pettini}, {Trainor}  \& {Strom}}{{Steidel}
  et~al.}{2018}]{Steidel2018}
{Steidel} C.~C.,  {Bogosavljevi{\'c}} M.,  {Shapley} A.~E.,  {Reddy} N.~A.,
  {Rudie} G.~C.,  {Pettini} M.,  {Trainor} R.~F.,   {Strom} A.~L.,  2018,
  \mn@doi [\apj] {10.3847/1538-4357/aaed28}, \href
  {https://ui.adsabs.harvard.edu/abs/2018ApJ...869..123S} {869, 123}

\bibitem[\protect\citeauthoryear{{Telfer}, {Zheng}, {Kriss}  \&
  {Davidsen}}{{Telfer} et~al.}{2002}]{Telfer2002}
{Telfer} R.~C.,  {Zheng} W.,  {Kriss} G.~A.,   {Davidsen} A.~F.,  2002, \mn@doi
  [\apj] {10.1086/324689}, \href
  {https://ui.adsabs.harvard.edu/abs/2002ApJ...565..773T} {565, 773}

\bibitem[\protect\citeauthoryear{{Trump} et~al.,}{{Trump}
  et~al.}{2009}]{Trump2009}
{Trump} J.~R.,  et~al., 2009, \mn@doi [\apj] {10.1088/0004-637X/696/2/1195},
  \href {http://adsabs.harvard.edu/abs/2009ApJ...696.1195T} {696, 1195}

\bibitem[\protect\citeauthoryear{{Vanden Berk} et~al.,}{{Vanden Berk}
  et~al.}{2001}]{VandenBerk2001}
{Vanden Berk} D.~E.,  et~al., 2001, \mn@doi [\aj] {10.1086/321167}, \href
  {https://ui.adsabs.harvard.edu/abs/2001AJ....122..549V} {122, 549}

\bibitem[\protect\citeauthoryear{{Vanden Berk}, {Wesolowski}, {Yeckley},
  {Marcinik}, {Quashnock}, {Machia}  \& {Wu}}{{Vanden Berk}
  et~al.}{2020}]{VandenBerk2020}
{Vanden Berk} D.~E.,  {Wesolowski} S.~C.,  {Yeckley} M.~J.,  {Marcinik} J.~M.,
  {Quashnock} J.~M.,  {Machia} L.~M.,   {Wu} J.,  2020, \mn@doi [\mnras]
  {10.1093/mnras/staa411}, \href
  {https://ui.adsabs.harvard.edu/abs/2020MNRAS.493.2745V} {493, 2745}

\bibitem[\protect\citeauthoryear{{Vito}, {Gilli}, {Vignali}, {Comastri},
  {Brusa}, {Cappelluti}  \& {Iwasawa}}{{Vito} et~al.}{2014}]{Vito2014}
{Vito} F.,  {Gilli} R.,  {Vignali} C.,  {Comastri} A.,  {Brusa} M.,
  {Cappelluti} N.,   {Iwasawa} K.,  2014, \mn@doi [\mnras]
  {10.1093/mnras/stu2004}, \href
  {http://adsabs.harvard.edu/abs/2014MNRAS.445.3557V} {445, 3557}

\bibitem[\protect\citeauthoryear{{Willott} et~al.,}{{Willott}
  et~al.}{2010}]{Willott2010}
{Willott} C.~J.,  et~al., 2010, \mn@doi [\aj] {10.1088/0004-6256/139/3/906},
  \href {http://adsabs.harvard.edu/abs/2010AJ....139..906W} {139, 906}

\makeatother
\end{thebibliography}





\appendix

\section{Magnitude error estimates}
\label{appendix:mag_error}

For HSC SSP data, error values from the database output
(\verb+[grizy]_psfflux_magerr+) are used as 1$\sigma$
errors PSF model-based  magnitude errors.
We compared the magnitude errors from the database with those computed
using \textsc{SExtractor} and confirmed that the errors by these two
independent software agree reasonably well.
For CLAUDS CFHT / MegaCam \fu and \fuS photometry, 
we use 1\farcs5-diameter aperture photometry to determine if an object is
detected or not. 
Aperture magnitude errors are 
estimated as follows. First for each patch where a target AGN
resides we generate a template point spread function by selecting bright
point sources in the patch and combine them after normalization (about
60 to 80 sources are used for each patch). Then 1,000 dummy point
sources with a spatial profile based on the template PSF with Poisson
noise fluctuation are 
added to the image. We measure their aperture counts with 1\farcs5
diameter aperture. The standard deviation of these values gives
1$\sigma$ error for point sources with a magnitude. This procedure is
repeated for magnitude range 21.0 to 28.5 with a 0.5 magnitude step for
each patch. The magnitude error of each AGN in our sample was determined
by interpolating results of these simulations.
For objects with $>3\sigma$ signal in aperture photometry, we use 
PSF model-based photometry for CLAUDS data in the HSC SSP database to
obtain flux ratios between $U$-band and $i$-band.


\bsp	
\label{lastpage}

\end{document}